\documentclass[12pt,letterpaper]{article}
\usepackage{epsfig,rotating,setspace,latexsym,amsmath,epsf,amssymb,amsfonts,bm,theorem,cite,caption,subcaption,enumerate,longtable,accents}
\usepackage{algorithm,algorithmic,graphicx,epsf,authblk,epstopdf,url,color,multirow}

\newtheorem{theorem}{Theorem}

\newtheorem{remark}{Remark}
\newtheorem{lemma}{Lemma}
\newtheorem{example}{Example}
\newenvironment{Proof}[1]{\medskip\par\noindent{\bf Proof:\,}\,#1}{{\mbox{\,$\blacksquare$}\par}}

\newcommand{\cp}{{\mathcal{P}}}

\allowdisplaybreaks

\begin{document}

\title{Private Set Intersection: A Multi-Message Symmetric Private Information Retrieval Perspective\thanks{This work was supported by NSF Grants CCF 17-13977 and ECCS 18-07348.}}

\author[1]{Zhusheng Wang}
\author[2]{Karim Banawan}
\author[1]{Sennur Ulukus}

\affil[1]{\normalsize Department of Electrical and Computer Engineering, University of Maryland}
\affil[2]{\normalsize Electrical Engineering Department, Faculty of Engineering, Alexandria University}

\maketitle

\vspace*{-0.5cm}

\begin{abstract}
We study the problem of private set intersection (PSI). In this problem, there are two entities $E_i$, for $i=1, 2$, each storing a set $\mathcal{P}_i$, whose elements are picked from a finite set $\mathbb{S}_K$, on $N_i$ replicated and non-colluding databases. It is required to determine the set intersection $\cp_1 \cap \cp_2$ without leaking any information about the remaining elements to the other entity, and to do this with the least amount of downloaded bits. We first show that the PSI problem can be recast as a multi-message symmetric private information retrieval (MM-SPIR) problem with certain added restrictions. Next, as a stand-alone result, we derive the information-theoretic sum capacity of MM-SPIR, $C_{MM-SPIR}$. We show that with $K$ messages, $N$ databases, and a given size of the desired message set $P$, the exact capacity of MM-SPIR is  $C_{MM-SPIR} = 1 - \frac{1}{N}$ when $P \leq K-1$, provided that the entropy of the common randomness $S$ satisfies $H(S) \geq \frac{P}{N-1}$ per desired symbol. When $P = K$, the MM-SPIR capacity is trivially $1$ without the need for any common randomness $S$. This result implies that there is no gain for MM-SPIR over successive single-message SPIR (SM-SPIR). For the MM-SPIR problem, we present a novel capacity-achieving scheme which builds seamlessly over the near-optimal scheme of Banawan-Ulukus originally proposed for the multi-message PIR (MM-PIR) problem without any database privacy constraints. Surprisingly, our scheme here is exactly optimal for the MM-SPIR problem for any $P$, in contrast to the scheme for the MM-PIR problem, which was proved only to be near-optimal. Our scheme is an alternative to the successive usage of the SM-SPIR scheme of Sun-Jafar. Based on this capacity result for the MM-SPIR problem, and after addressing the added requirements in its conversion to the PSI problem, we show that the optimal download cost for the PSI problem is given by $\min\left\{\left\lceil\frac{P_1 N_2}{N_2-1}\right\rceil, \left\lceil\frac{P_2 N_1}{N_1-1}\right\rceil\right\}$, where $P_i$ is the cardinality of set $\cp_i$.
\end{abstract}

\section{Introduction}
The private set intersection (PSI) problem refers to the problem of determining the common elements in two sets (lists) without leaking any further information about the remaining elements in the sets. This problem has been a major research topic in the field of cryptography starting with the work \cite{PSI_first} (see also \cite{PSI_survey, PSI_computational,PSI_efficient}). In all these works, computational guarantees are used to ensure the privacy of the elements beyond the intersection. The PSI problem can be motivated by many practical examples, for instance: Consider an airline company which has a list of its customers, and a law enforcement agency which has a list of suspected terrorists. The airline company and the law enforcement agency wish to determine the intersection of their respective lists without the airline company revealing the rest of its customers and the law enforcement agency revealing the rest of the suspects in its list. As another example, consider a major service provider (e.g., Whatsapp) and a new customer who wishes to join this service. The user wishes to find out which members of his/her contact list are already using this service without revealing his/her entire contact list to the service provider. Similarly, the service provider wishes to determine the intersection without revealing its entire list of customers. For other examples, please see \cite{PSI_computational, PSI_survey}.  

Since the entities in PSI want to \emph{privately retrieve} the elements that belong to the intersection of their sets $\cp_1 \cap \cp_2$, where $\cp_i$ is the set (list) that belongs to the $i$th entity, private information retrieval (PIR) can be used as a building block for the PSI problem. In classical PIR, which was introduced by Chor et al. \cite{PIR_ORI}, a user wants to retrieve a message (file) from distributed databases without leaking any information about the identity of the desired file. This is desirable in the PSI problem, as one of the entities wants to retrieve the intersection $\cp_1 \cap \cp_2$. Nevertheless, it is needed to keep the remaining elements of the sets secret from the other entity, i.e., the first entity wants to keep the set $\cp_1 \setminus \cp_2$ from the second entity and vice versa. This gives rise naturally to the problem of symmetric PIR (SPIR), which was originally introduced in \cite{SPIR_ORI}, where the retrieval scheme needs to ensure that the user learns no information beyond the desired message. This extra requirement is called the \emph{database privacy} constraint, which is in addition to the usual \emph{user privacy} constraint in PIR. Recently, Sun and Jafar reformulated the problems of PIR and SPIR from an information-theoretic point of view, and determined the fundamental limits of both of these problems, i.e., their capacity, in \cite{PIR} and \cite{SPIR}, respectively. Subsequently, the fundamental limits of many interesting variants of PIR and SPIR have been considered, see for example \cite{JafarColluding, arbitraryCollusion, RobustPIR_Razane, Staircase_PIR, codedsymmetric, wang2017linear, SPIR_Mismatched, symmetricByzantine,  ChaoTian_leakage, KarimCoded, codedcolluded, codedcolludingZhang, Kumar_PIRarbCoded, codedcolludingJafar, MM-PIR, MPIRcodedcolludingZhang, BPIRjournal, CodeColludeByzantinePIR, tandon2017capacity, KimCache, wei2017fundamental, wei2017fundamental_partial, PIR_cache_edge, kadhe2017private, chen2017capacity, wei2017capacity, MMPIR_PSI, SSMMPIR_SI1, SSMMPIR_SI2, LiConverse, StorageConstrainedPIR_Wei, PrivateComputation, mirmohseni2017private, PrivateSearch, abdul2017private, StorageConstrainedPIR, efficient_storage_ITW2019, Chao_storage_cost, PIR_decentralized, heteroPIR, TamoISIT, Karim_nonreplicated, PIR_WTC_II, SecurePIR, securePIRcapacity, securestoragePIR,  XSTPIR, arbmsgPIR, ChaoTian_coded_minsize, MultiroundPIR, KarimAsymmetricPIR, noisyPIR, PIR_lifting, PIR_networks}. 

Now, to use SPIR to implement PSI, the $i$th entity needs to privately check the presence of each element in $\cp_i$ at the other entity. That is, the $i$th entity needs to retrieve the \emph{occurrences} of all elements that belong to its set $\cp_i$ from the other entity. This implies that the $i$th entity needs to retrieve \emph{multiple messages} from the other entity, where the messages here correspond to the \emph{incidences} of each element of the set $\cp_i$. This establishes the connection between the PSI problem and the \emph{multi-message} SPIR (MM-SPIR) problem. Apart from the PSI problem, the MM-SPIR problem is interesting on its own right and has remained an open problem until this work. Reference \cite{MM-PIR} investigates the problem of multi-message PIR (MM-PIR) without any database privacy constraints. The results of \cite{MM-PIR} show that the user can improve the retrieval rate by jointly retrieving the desired messages instead of retrieving them one-by-one. In this paper, we aim to characterize the capacity of the MM-SPIR problem as a stand-alone result, and determine whether the MM-SPIR capacity is larger than the single-message SPIR (SM-SPIR) capacity. Second, we aim to unify the achievability schemes of MM-PIR and MM-SPIR so that the query structure can be maintained with and without the database privacy constraints.  

The papers that are most closely related to our work are the ones that focus on \emph{symmetry} and \emph{multi-message} aspects of PIR. Reference \cite{SPIR} derives the SPIR capacity when the user wishes to retrieve a single message as $C_{SM-SPIR}=1-\frac{1}{N}$. Reference \cite{MM-PIR} considers MM-PIR and determines the exact capacity when the number of desired messages $P$ is at least half of the total number of messages $K$ or when $K/P$ is an integer; for all other cases \cite{MM-PIR} provides a novel PIR scheme which is near-optimal. Reference \cite{MMPIR_PSI} studies multi-server MM-PIR with private side information. References \cite{SSMMPIR_SI1, SSMMPIR_SI2} study single-server MM-PIR with side information. Reference \cite{codedsymmetric} studies SPIR from MDS-coded databases. The problem is extended to include colluding servers in \cite{wang2017linear} and mismatches between message and common randomness codes in \cite{SPIR_Mismatched}. Reference \cite{symmetricByzantine} investigates SPIR in the presence of adversaries. Reference \cite{ChaoTian_leakage} characterizes the tradeoff between the minimum download cost and the information leakage from undesired messages. None of these works considers the interplay between the data privacy constraint and the joint retrieval of multiple messages, as needed in MM-SPIR. 
     
In this paper, first focusing on MM-SPIR as a stand-alone problem, we derive its capacity. Our results show that the sum capacity of MM-SPIR is exactly equal to the capacity of SM-SPIR, i.e., $C_{SM-PIR}=C_{MM-PIR}=1-\frac{1}{N}$. We show that the databases need to share a random variable $S$ such that $H(S) \geq \frac{P}{N-1}$ per desired symbol, which is $P$ multiple of the common randomness required for SM-SPIR. This implies that, unlike MM-PIR, there is no gain from jointly retrieving the $P$ messages, and it suffices to download the $P$ messages successively using the SM-SPIR scheme in \cite{SPIR}, provided that statistically independent common randomness symbols are used at each time. For the extreme case $P=K$, i.e., when the user wants to retrieve all messages, the problem reduces to SPIR  with $K=1$ message, where the database privacy and the user privacy constraints are trivially satisfied and full capacity (i.e., $C_{MM-SPIR}=1$) is attained without the need for any common randomness.

Further, for MM-SPIR, we propose a novel capacity-achieving scheme for $1 \leq P \leq K-1$. Compared with the one in \cite{SPIR}, the form of this achievable scheme is much closer to the achievable scheme in \cite{PIR}. The query structure of the scheme resembles its counterpart in \cite{MM-PIR}, in particular, we construct the greedy algorithm in \cite{PIR} backwards as in \cite{MM-PIR}. The major difference between our proposed scheme here and the MM-PIR scheme in \cite{MM-PIR} is the fact that databases add the common randomness to the returned answer strings to satisfy the database privacy constraint. Our scheme is surprisingly optimal for all $P$ and $K$ in contrast to the scheme in \cite{MM-PIR} which is proved to be optimal only if $P$ is at least half of $K$ or $K/P$ is an integer. By plugging $P=1$, our scheme serves as an alternative capacity-achieving scheme for the SM-SPIR scheme in \cite{SPIR}. As an added advantage, our scheme extends seamlessly the MM-PIR scheme to satisfy the database privacy constraint without changing the query structure. Hence, by operating such a scheme the databases can support SPIR and PIR simultaneously. Moreover, the scheme may serve as a stepping stone to solve some other SPIR problems, such as, SM-SPIR or MM-SPIR with side information.
     
In this paper, we ultimately consider the PSI problem. There are two entities $E_1$ and $E_2$. The entity $E_i$ has a set (list) $\cp_i$, whose elements are picked from a finite set $\mathbb{S}_K$ and has a cardinality $P_i$. The set $\cp_i$ is stored on $N_i$ non-colluding and replicated databases. It is required to compute the intersection $\cp_1 \cap \cp_2$ without leaking information about $\cp_1 \setminus \cp_2$ or $\cp_2 \setminus \cp_1$ with the minimum download cost. We first show that this problem can be recast as an MM-SPIR problem, where a user needs to retrieve $P$ messages from a library containing $K$ messages. In this MM-SPIR problem, messages correspond to \emph{incidences} of elements in these sets with respect to the field elements. Specifically, the entity $E_i$ constructs the incidence vector of its elements with respect to the field elements. The incidence vector is a binary vector of length $K$ that stores a $1$ in the position of the $j$th element of the field if this field element is in $\cp_i$. This transforms each set into a library of $K$ binary messages (of length $1$ bit each). This transformation is needed since in SPIR, a user needs to know the location of the file(s) in the databases. Therefore, in transforming the PSI problem into an MM-SPIR problem, two restrictions arise: First, the message size is fixed and finite, which is $1$ in this case. Second, depending on the model assumed regarding the generation of sets $\cp_1$ and $\cp_2$, the messages may be correlated. In our formulation, the message size is 1, but the messages are independent; see the exact problem formulation. Following these constructions, entity $E_i$ performs MM-SPIR of the messages corresponding to its set $\cp_i$ within the databases of the other entity. By decoding these messages, the intersection $\cp_1 \cap \cp_2$ is determined without leaking any information about $\cp_1 \setminus \cp_2$ or $\cp_2 \setminus \cp_1$. This is a direct consequence of satisfying the reliability, user privacy, and database privacy constraints of the MM-SPIR problem. We show that the optimum download cost of the PSI problem is $\min\left\{\left\lceil\frac{P_1 N_2}{N_2-1}\right\rceil, \left\lceil\frac{P_2 N_1}{N_1-1}\right\rceil\right\}$, which is linear in the size of the smaller set, i.e., $\min\{P_1, P_2\}$. The linear scaling appears in the problem of determining the set intersection even without any privacy constraints. 

\section{PSI: Problem Formulation} \label{PSI-formulation}
Consider the problem of privately determining the intersection of two sets (or lists) picked from a finite set\footnote{The restriction of generating the set from a finite set is without loss of generality as the set elements of any kind can be mapped into corresponding finite set elements for sufficiently large size. For example, the elements of the set that contains the names of suspected terrorists in the United States can be mapped into elements from the finite set $\mathbb{S}_K$, where $K$ is the population size on this planet. As we will show next, the download cost is independent of $K$. Hence, the optimization of the alphabet size is irrelevant to our formulation. Nevertheless, it is advisable to choose $K$ to be the lowest integer such that $\cp_1,\cp_2 \subseteq \mathbb{S}_K$ to minimize the upload cost. It suffices to have $K>P_1+P_2$.} $\mathbb{S}_K$. For convenience, we denote a random variable and its realization by using the same general uppercase letter when distinction is clear from the context. We address this issue additionally whenever clarification is needed. Consider a setting where there are two entities $E_1$ and $E_2$. For $i=1,2$, the entity $E_i$ stores a set $\cp_i$. For each element of the finite set $\mathbb{S}_K$, the entity $E_i$ adds\footnote{\label{footnote2}We note that our achievability scheme works for any statistical distribution imposed on the sets, i.e., the i.i.d.~generation assumption presented here is not needed for the achievability proof.} this element to its set $\cp_i$ independently from the remaining field elements with probability $q_i$. In this work, we focus on the case of $q_i=\frac{1}{2}$ for $i=1,2$. After generation of the set $\cp_i$, the cardinality of $\cp_i \subseteq \mathbb{S}_K^{P_i}$ is denoted by $|\cp_i|=P_i$, and is public knowledge.\footnote{We note that choosing to have $P_i$ to be a global knowledge is for convenience only. This knowledge enables the entities to determine which entity should initiate the PSI process to have the least download cost (or if any is needed at all, as in the case of $P_i=K$, for an $i$; see Remark~\ref{remark1}). If the cardinalities are not public knowledge, our achievability works by choosing one of the entities arbitrarily to initiate the PSI process assuming that the other entity has sufficient common randomness. We note, however, that keeping the cardinalities private is indeed a challenging problem and it is outside the scope of this work.} The entity $E_i$ stores $\cp_i$ in a replicated fashion on $N_i$ replicated and non-colluding databases.

\begin{figure}[t]
	\centering
	\includegraphics[width=\textwidth]{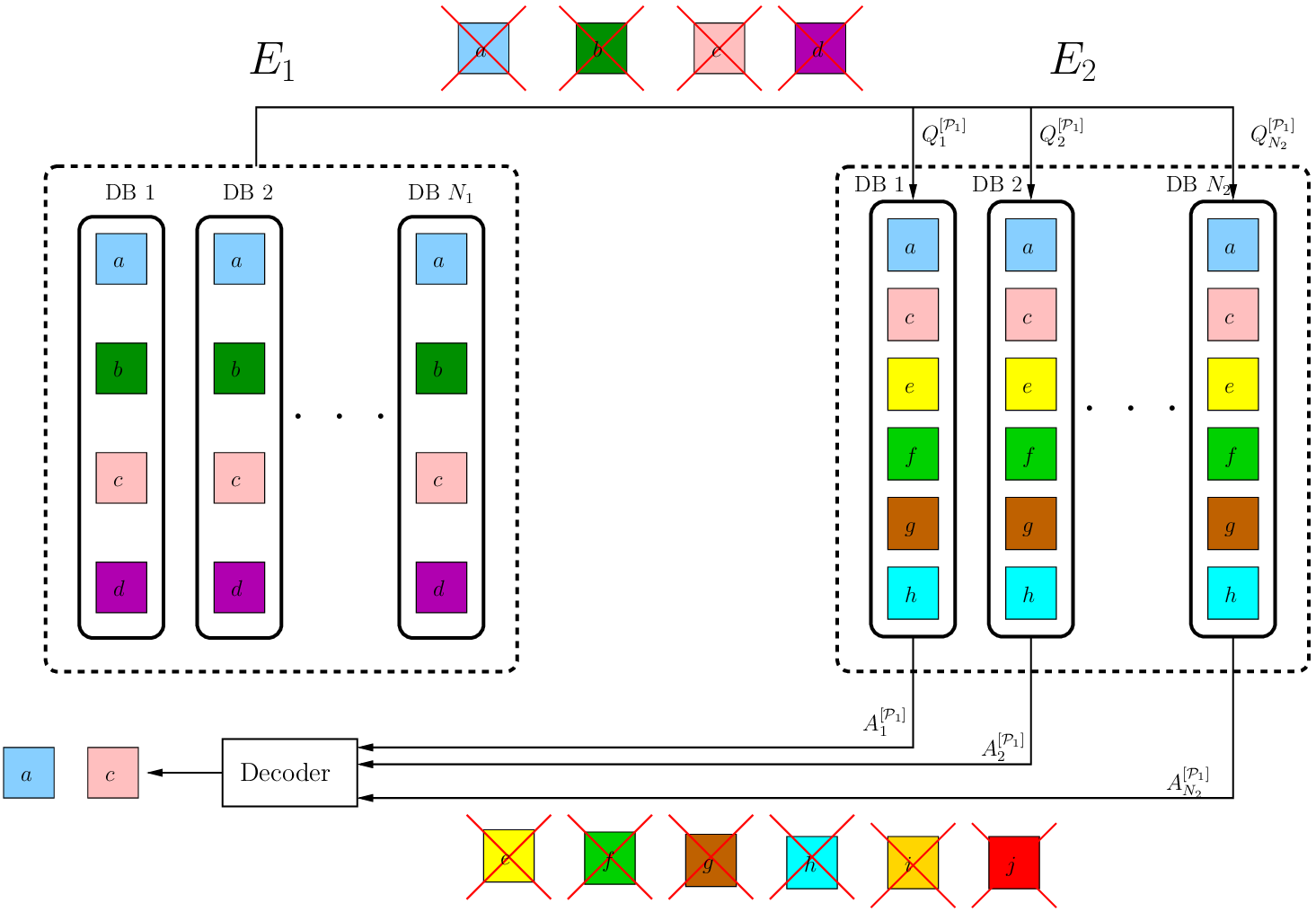}
	\caption{Example for the private set intersection (PSI) problem. $E_1$ has the set $\mathcal{P}_1=\{a,b,c,d\}$ and $E_2$ has the set $\mathcal{P}_2=\{a,c,e,f,g,h\}$. $E_1$ submits queries to $E_2$ that do not leak information about $\cp_1$, while $E_2$ responds with answers that do not leak information about $e,f,g,h$ (or non-existence of $i,j$). By decoding the answers, $E_1$ learns that $\mathcal{P}_1 \cap \mathcal{P}_2=\{a,c\}$.}
	\label{PSI_original}
\end{figure}

The entities $E_1$ and $E_2$ want to compute the intersection $\cp_1 \cap \cp_2$ privately (see Fig.~\ref{PSI_original}). To that end, the entity\footnote{The entities $E_1$, $E_2$ should agree on a specific order of retrieval operations such that this order results in the minimal download cost. Without loss of generality, we assume here that the optimal order of operation starts with entity $E_1$ sending queries to the databases associated with entity $E_2$.} $E_1$ sends $N_2$ queries to the databases associated with $E_2$. Specifically, $E_1$ sends the query $Q_{n_2}^{[\cp_1]}$ to the $n_2$th database for all $n_2 \in [N_2]$, where $[N_2]$ (and also $[1:N_2]$) denotes integers from $1$ to $N_2$. Since $E_1$ does not know $\cp_2$ in advance, it generates the queries $Q_{1:N_2}^{[\cp_1]}=\left\{Q_{n_2}^{[\cp_1]}: n_2 \in [N_2]\right\}$ independently from $\cp_2$, hence,
\begin{align}\label{indep}
I(Q_{1:N_2}^{[\cp_1]};\cp_2)=0
\end{align}

\sloppy The databases associated with $E_2$ respond truthfully with answers $A_{1:N_2}^{[\cp_1]}=\left\{A_{n_2}^{[\cp_1]}: n_2 \in [N_2]\right\}$. The $n_2$th answer $A_{n_2}^{[\cp_1]}$ is a deterministic function of the set $\cp_2$, the query $Q_{n_2}^{[\cp_1]}$ and the existing common randomness $S$, thus,
\begin{align}
H(A_{n_2}^{[\cp_1]}|Q_{n_2}^{[\cp_1]},\cp_2,S)=0, \quad n_2 \in [N_2]
\end{align}

Denote the cardinality of the intersection $|\cp_1 \cap \cp_2|$ by $M$. The entity\footnote{After calculating $\cp_1 \cap \cp_2$ at $E_1$, the entity $E_1$ sends the result of $\cp_1 \cap \cp_2$ directly to $E_2$ if needed.} $E_1$ should be able to reliably compute the intersection $\cp_1 \cap \cp_2$ based on the sent queries $Q_{1:N_2}^{[\cp_1]}$, the collected answers $A_{1:N_2}^{[\cp_1]}$ and the knowledge of $\mathcal{P}_1$ without knowing $M$ in advance. This is captured by the following PSI reliability constraint,
\begin{align}\label{PSI Reliability}
\text{[PSI reliability]} \qquad  H(\cp_1 \cap \cp_2|Q_{1:N_2}^{[\cp_1]}, A_{1:N_2}^{[\cp_1]},\mathcal{P}_1)=0
\end{align}

The privacy requirements can be expressed as the following two privacy constraints: $E_1$ privacy and $E_2$ privacy. First, the queries sent by $E_1$ should not leak any information about\footnote{While checking the presence of elements of $\cp_1$ in $\cp_2$, $E_1$ wants to protect $\cp_1\setminus \cp_2$. However, since $E_1$ does not know $\cp_2$, the queries cannot depend on $\cp_2$ (see also \eqref{indep}), and $E_1$ should protect all of $\cp_1$ in queries.} $\cp_1$, i.e., any individual database associated with $E_2$ learns nothing about $\cp_1$ from the query $Q_{n_2}^{[\cp_1]}$, the answer $A_{n_2}^{[\cp_1]}$, the knowledge of $\mathcal{P}_2$ and the existing common randomness $S$, 
\begin{align}\label{E1 privacy} 
\text{[$E_1$ privacy]} \qquad I(\cp_1;Q_{n_2}^{[\cp_1]},A_{n_2}^{[\cp_1]},\mathcal{P}_2,S)=0, \quad n_2 \in [N_2]
\end{align}
Second, $E_1$ should not be able to learn anything further than $\cp_1 \cap \cp_2$, i.e., $E_1$ should not learn the elements in $\cp_2$ other than the intersection, $\cp_2\setminus (\cp_1 \cap \cp_2) = \cp_2\setminus \cp_1$. Moreover\footnote{Although it is tempting to formulate the $E_2$ privacy constraint as $I(\cp_2 \setminus \cp_1;A_{1:N_2}^{[\cp_1]})=0$, this constraint permits leaking information about the remaining field elements that do not exist in $\cp_2$. More specifically, if we adopted this constraint in the example in Fig.~\ref{PSI_original}, the answers should not leak information about $e,f,g,h$, however, $E_1$ may learn that the elements $i,j$ do not exist in $\cp_2$. To properly formalize the constraint that $E_1$ learns nothing other than the intersection, we need to protect $\overline{(\cp_1 \cup \cp_2)}$ as well.}, $E_1$ should not learn the absence of the remaining field elements in $E_2$, i.e., the set   $\overline{(\cp_1 \cup \cp_2)}$. Thus, $E_1$ should learn nothing about whether $E_2$ contains $(\cp_2\setminus \cp_1) \cup \overline{(\cp_1 \cup \cp_2)}=\bar{\cp}_1$ or not (we denote this information by $E_{2,\bar{\cp}_1}$) from the collected answers $A_{1:N_2}^{[\mathcal{P}_1]}$ given the generated queries $Q_{1:N_2}^{[\mathcal{P}_1]}$ and the knowledge of $\mathcal{P}_1$,
\begin{align}\label{E2 privacy} 
\text{[$E_2$ privacy]} \qquad  I(E_{2,\bar{\cp}_1};Q_{1:N_2}^{[\mathcal{P}_1]},A_{1:N_2}^{[\mathcal{P}_1]}, \mathcal{P}_1)=0
\end{align}

For given finite set size $K$, set sizes $P_1$ and $P_2$, and number of databases $N_1$ and $N_2$, an achievable PSI scheme is a scheme that satisfies the PSI reliability constraint \eqref{PSI Reliability}, the $E_1$ privacy constraint \eqref{E1 privacy}, and the $E_2$ privacy constraint \eqref{E2 privacy}. In this paper, we measure the efficiency of a scheme by the maximal number of downloaded bits by one of the entities $E_1$ or $E_2$ in order to compute $\cp_1 \cap \cp_2$. We denote the maximal number of downloaded bits by $D$. Then, the optimal download cost is $D^*=\inf D$ over all achievable PSI schemes.\footnote{A more natural efficiency metric is to consider the sum of the maximal number of uploaded bits (denoted by $U$) and the maximal number of downloaded bits (denoted by $D$) by one of the entities $E_1$ or $E_2$ to compute $\cp_1 \cap \cp_2$. In this case, the most efficient scheme is the scheme with the lowest communication cost, i.e., that achieves the optimal communication cost $C^*=\inf (U+D)$ over all achievable PSI schemes. The SPIR problem \cite{SPIR} under combined upload and download costs is still an open problem. As we will see, our framework builds on the SPIR problem. Therefore, in this work, we consider only the download cost. The PSI under combined upload and download costs is an interesting future direction, which is outside the scope of our paper. In Section~\ref{upload}, we provide an illustrative example to show that the upload cost can be reduced without affecting the download cost. Nevertheless, we argue that if the PSI determination is repeated (for example, if one list is kept the same and the other list is regularly updated, we always use the fixed list to initiate the PSI process), the queries could be used repeatedly without compromising the user privacy as long as the databases do not collude. In this case, the upload cost would not scale with the number of PSI determination rounds, unlike the download cost.}
     
\section{From PSI to MM-SPIR} \label{PSI-transformation}
\begin{figure}[t]
	\centering
	\includegraphics[width=0.8\textwidth]{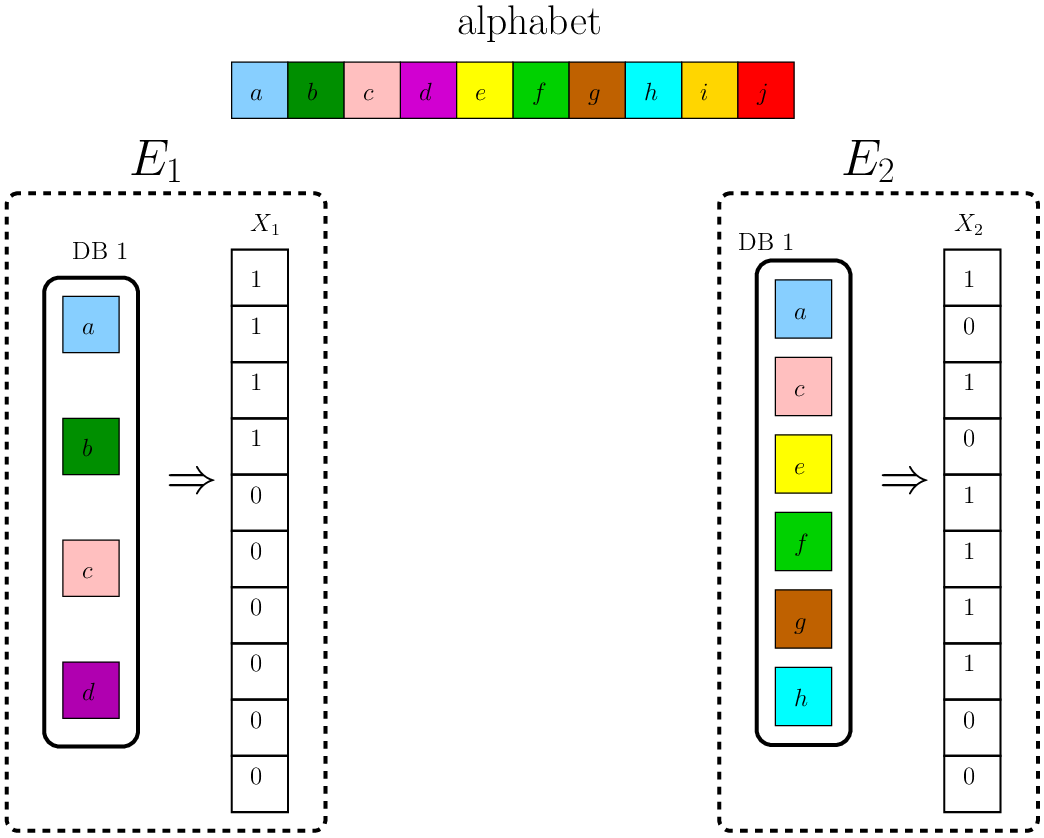}
	\caption{Example for the transformation from sets to incidence vectors. $E_1$ has the set $\cp_1=\{a,b,c,d\}$ and $E_2$ has the set $\cp_2=\{a,c,e,f,g,h\}$. The alphabet is $\cp_{alph}=\{a, b, c, d, e, f, g, h, i, j\}$. Entity $E_i$ constructs an incidence vector $X_i$ to facilitate MM-SPIR.}
	\label{PSI_transformation}
\end{figure}

\begin{figure}[t]
	\centering
	\includegraphics[width=\textwidth]{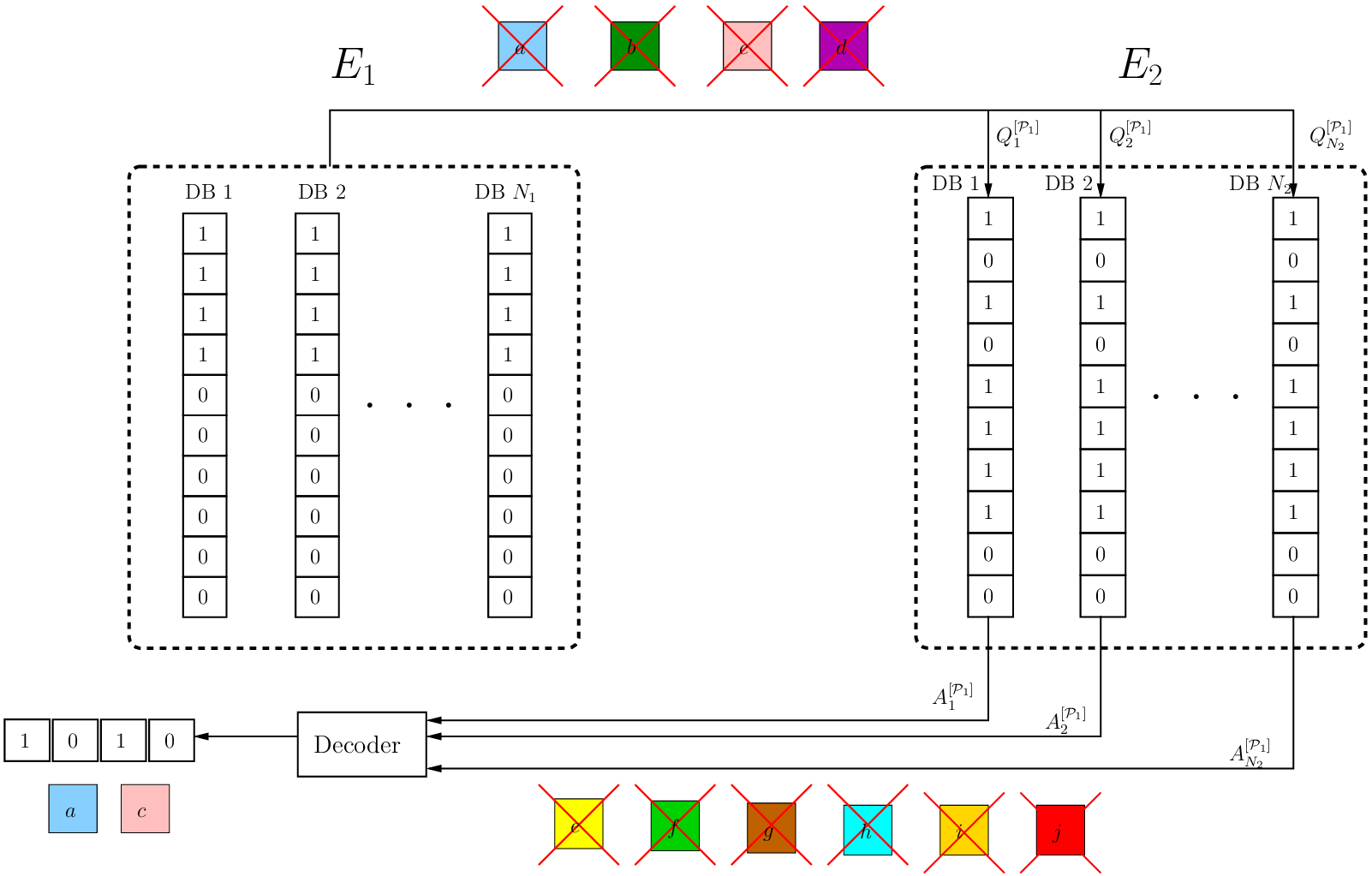}
	\caption{Example for the transformation from the PSI problem to an MM-SPIR problem. $E_1$ needs to retrieve the elements corresponding to $\cp_1$ from the incidence vector $X_2$ without revealing $\cp_1$, while $E_2$ responds with answer strings that do not leak $\bar{\cp}_1$.}
	\label{PSI_SPIR}
\end{figure}

In this section, we show that the PSI problem can be reduced to an MM-SPIR problem, if the entities allow storing their sets in a specific searchable format. This transformation has the same flavor as \cite{PIRkeywords} and \cite{PrivateSearch}, where the original contents of the databases are mapped into searchable lists to enable PIR, which assumes that the user knows the position of the desired file in the databases. To that end, define the incidence vector $X_i \in \mathbb{F}_2^K$ as a binary vector of size $K$ associated with the set $\cp_i$. Denote the $j$th element of the incidence vector $X_i$ by $X_i(j)$ where
\begin{align}
X_i(j)=\begin{cases}
1, \quad j \in \cp_i \\
0, \quad j \notin \cp_i
\end{cases}
\end{align}
for all $j \in \mathbb{S}_K$. Hence, $X_i(j)$ is an i.i.d.~random variable for all $j \in [K]$ such that $X_i(j) \sim \text{Ber}(q_i)$. The entity $E_i$ constructs the incidence vector $X_i$ corresponding to the set $\cp_i$ (see Fig.~\ref{PSI_transformation}). The entity $E_i$ replicates the vector $X_i$ at all of its $N_i$ associated databases (see Fig.~\ref{PSI_SPIR}). Note that $X_i$ is a sufficient statistic for $\cp_i$ for a given $K$. The PSI determination process is performed over $X_1$ or $X_2$, and not over the original $\cp_1$ or $\cp_2$. 

To solidify ideas, we state the variables defined so far explicitly over a specific example. Consider the example in Fig.~\ref{PSI_original}. Here, the entity $E_1$ has the set $\cp_1=\{a, b, c, d\}$ and the entity $E_2$ has the set $\cp_2=\{a, c, e, f, g, h\}$. Therefore, the intersection is $\cp_1 \cap \cp_2=\{a, c\}$. Let us assume that the alphabet, $\cp_{alph}$, for this example is $\cp_{alph}=\{a, b, c, d, e, f, g, h, i, j\}$ as shown in Fig.~\ref{PSI_transformation}. Then, the incidence vectors at the entities are $X_1=[1 ~ 1 ~ 1 ~ 1 ~ 0 ~ 0 ~ 0 ~ 0 ~ 0 ~ 0]$ and $X_2=[1 ~ 0 ~ 1 ~ 0 ~ 1 ~ 1 ~ 1 ~ 1 ~ 0 ~ 0]$, which are also shown in Fig.~\ref{PSI_transformation}. For this example, $P_1=4$, $P_2=6$, $K=10$, and $M=2$. Finally, the MM-SPIR is conducted over the replicated incidence vectors at the two entities as shown in Fig.~\ref{PSI_SPIR}.

Without loss of generality, assume that $E_1$ initiates the PSI process. $E_1$ does not know $M$ in advance. The only information $E_1$ has is $\cp_1$. Consequently, $E_1$ wants to verify the existence of each element of $\cp_1$ in $\cp_2$ to deduce $\cp_1 \cap \cp_2$. Thus, $E_1$ needs to jointly and reliably download the bits $W_{\mathcal{P}_1}=\{X_2(j): j \in \cp_1\}$ by sending $N_2$ queries to the databases associated with $E_2$ and collecting the corresponding answers with the knowledge of $X_1$. Hence, we can write the reliability constraint as,
\begin{align}
 H(W_{\mathcal{P}_1}|Q_{1:N_2}^{[\cp_1]}, A_{1:N_2}^{[\cp_1]}, X_1)=0
\end{align}

Since $E_1$ is searching for the existence of all elements of $\cp_1$ in $\cp_2$ without leaking any information about $\cp_1$ to any individual database associated with $E_2$, the $E_1$ privacy constraint in \eqref{E1 privacy} dictates,
\begin{align}
I(\cp_1;Q_{n_2}^{[\cp_1]},A_{n_2}^{[\cp_1]},X_2,S)=0, \quad n_2 \in [N_2] \label{modified-E1-equiv}
\end{align}
This is exactly the privacy constraint in the MM-PIR problem \cite{MM-PIR}. 

As the databases associated with $E_2$ store $X_2$ now, to ensure the $E_2$ privacy constraint in \eqref{E2 privacy}, the answers from $E_2$ databases should not leak anything about $E_{2,\bar{\cp}_1}$, which can be further mapped to not leaking any information about $W_{\bar{\cp}_1}=\{X_2(j): j \notin \cp_1\}$ as, 
\begin{align}
 I(W_{\bar{\mathcal{P}}_1};Q_{1:N_2}^{[\mathcal{P}_1]},A_{1:N_2}^{[\mathcal{P}_1]}, X_1) = 0
\end{align}
This is exactly the database privacy constraint in MM-SPIR; see Section~\ref{MM-SPIR}.

Consequently, the PSI problem formally reduces to MM-SPIR with i.i.d.~messages of length $1$ bit each (see Fig.~\ref{PSI_SPIR}), when the entities $E_1$ and $E_2$ are allowed to construct the corresponding incidence vectors for the original sets $\cp_1$ and $\cp_2$. The message length constraint of $1$ bit per message, i.e., $H(W_k)=1$ for all $k \in [K]$, comes due to messages representing incidences in the SPIR problem. The i.i.d.~property of the messages that we have here in this paper is a consequence of the i.i.d.~generation of the sets with probability $q_i$, and it is not true in general. In Section~\ref{MM-SPIR section}, we derive in detail the capacity of the MM-SPIR problem (see also Section~\ref{MM-LSPIR}), which in turn  gives the most efficient information-theoretic PSI scheme.

\section{Main Result}
In this section, we present our main result concerning the PSI problem. The result provides the optimal (minimum) download cost for the PSI problem under the assumptions in Sections~\ref{PSI-formulation} and \ref{PSI-transformation}. The result is based on the optimal download cost of the MM-SPIR problem, which is presented in detail in Section~\ref{MM-SPIR section}; see also Section~\ref{MM-LSPIR}.

\begin{theorem} \label{Thm1}
	 In the PSI problem, the elements of the sets are added independently with probability $q_i=\frac{1}{2}$ from a finite set of size $K$. Once the set generation is finished, the fixed set $\cp_1$ where $|\cp_1| = P_1 < K$ is stored among $N_1$ databases and the fixed set $\cp_2$ where $|\cp_2| = P_2 < K$ is stored among $N_2$ databases. The set cardinalities $P_1$ and $P_2$ are made public. The amount of common randomness satiesfies $H(S) \geq \min{\{\left\lceil\frac{P_1}{N_2-1}\right\rceil,\left\lceil\frac{P_2}{N_1-1}\right\rceil\}}$. Then, the optimal download cost with one-round communication (one entity sends the queries to the other entity and then receives feedback) is,
	\begin{align}
	D^*=\min\left\{\left\lceil\frac{P_1 N_2}{N_2-1}\right\rceil, \left\lceil\frac{P_2 N_1}{N_1-1}\right\rceil\right\} 
	\end{align}
\end{theorem}

The proof of Theorem~\ref{Thm1} is a direct consequence of the capacity result for MM-SPIR presented in Section~\ref{MM-SPIR section}; see also Section~\ref{MM-LSPIR}. We have the following remarks.

\begin{remark} \label{remark1}
    In the special case of having $P_i=K$ for $i=1$ or $i=2$, the download cost is trivially zero. This is due to the fact that if $P_1=K$ for example, the entity $E_2$ directly concludes that the intersection $\cp_1 \cap \cp_2=\cp_2$ without sending any queries to $E_1$ or requiring any common randomness.
\end{remark}

\begin{remark}
	The $\min$ term in Theorem~\ref{Thm1} comes from the fact that either entity can initiate the PSI determination process so that the overall download cost is minimized.
\end{remark}

\begin{remark}
    We note that although our result is exact, i.e.,  the download cost capacity (in the sense of matching achievability and converse proofs) under the assumptions of independent generation model for the lists with $q_i=\frac{1}{2}$, our scheme is achievable for any list generation model with arbitrary $q_i$ (see Footnote~\ref{footnote2}).
\end{remark}
	
\begin{remark}
	 Our result is private in information-theoretic (absolute) sense and does not need any assumptions about the computational powers of the entities. Furthermore, the achievable scheme is fairly simple and easy to implement compared to the fully homomorphic encryption needed in \cite{PSI_computational}.  A drawback of our approach is that it needs multiple non-colluding databases ($N_1$ or $N_2$ needs to be strictly larger than 1), otherwise, our scheme is infeasible.
\end{remark}

\begin{remark}
	The linear scalability of our scheme matches the linear scalability of the best-known set intersection algorithms without any privacy constraints. 
\end{remark}

\section{MM-SPIR as a Stand-Alone Problem}\label{MM-SPIR section}
In this section, we consider the MM-SPIR problem. We present the problem in a stand-alone format, i.e., we present a formal problem description in Section~\ref{Formulation}, followed by the main result in Section~\ref{Results}, the converse in Section~\ref{converse}, and a novel achievability in Section~\ref{achievability}. 

\subsection{MM-SPIR: Formal Problem Description}\label{MM-SPIR} \label{Formulation}
There are $N$ non-colluding databases each storing $K$ i.i.d.~messages. Each message is composed of $L$ \footnote{As in most PIR problems, the message length $L$ can approach infinity.} i.i.d.~and uniformly chosen symbols from a sufficiently large finite field $\mathbb{F}_q$. Then,
\begin{align}
H(W_{k}) &= L, \quad k \in [K]\\
H(W_{1:K}) &= KL
\end{align}

In the MM-SPIR problem, our goal is to retrieve a set of messages $W_{\cp}$ out of the $K$ available messages without leaking any information regarding the index set $\mathcal{P}$ to any individual database where $\cp =\{i_1,i_2,\cdots,i_P\} \subseteq [K]$ such that its cardinality is $|\cp| = P$.\footnote{We use the symbol $\cp$ to denote the random variable corresponding to the desired set and its realization with little abuse of notation.} This is the user privacy constraint. In addition, our goal is to not retrieve any messages beyond the desired set of messages $W_{\cp}$. This is the database privacy constraint.

Following the SPIR formulation in \cite{SPIR}, let $\mathcal{F}$ denote the randomness in the retrieving strategy adopted by the user. Because of the user privacy constraint, $\mathcal{F}$ is a random variable whose realization is only known to the user, but is unknown to the databases. A necessary common randomness $S$ must be shared among the $N$ databases to satisfy the database privacy constraint. The random variable $S$ is generated independent of the message set $W_{1:K}$. Similarly, $\mathcal{F}$ is independent of $W_{1:K}$ as the user does not know message realizations in advance. Moreover, $\mathcal{F}$ and $S$ are generated independently without knowing the desired index set $\mathcal{P}$. Then,
\begin{align}
H(\mathcal{F},S,\mathcal{P},W_{1:K}) = H(\mathcal{F}) + H(S) + H(\mathcal{P}) + H(W_{1:K}) \label{all independent}
\end{align}

To perform MM-SPIR, a user generates one query $Q_n^{[\mathcal{P}]}$ for each database according to the randomness $\mathcal{F}$ and then sends it to the $n$th database. Hence, the queries $Q_{1:N}^{[\mathcal{P}]}$ are deterministic functions of $\mathcal{F}$, i.e.,
\begin{align}
H(Q_1^{[\mathcal{P}]},Q_2^{[\mathcal{P}]},\cdots,Q_N^{[\mathcal{P}]}|\mathcal{F}) = 0, \quad \forall \mathcal{P} 
\label{random strategy}
\end{align}
Combining \eqref{all independent} and \eqref{random strategy}, the queries are independent of the messages, i.e.,
\begin{align}
I(Q_{1:N}^{[\mathcal{P}]};W_{1:K}) = 0 \label{queries and messages}   
\end{align}

After receiving a query from the user, each database truthfully generates an answer string based on the messages and the common randomness, hence,
\begin{align}
H(A_n^{[\mathcal{P}]}|Q_n^{[\mathcal{P}]},W_{1:K},S) = 0, \quad \forall n, \forall \mathcal{P} \label{determined answer string}
\end{align}

After collecting all the answer strings from the $N$ databases, the user should be able to decode the desired messages $W_{\mathcal{P}}$ reliably, therefore,
\begin{align}
\text{[reliability]} \quad H(W_{\mathcal{P}}|A_{1:N}^{[\mathcal{P}]},Q_{1:N}^{[\mathcal{P}]},\mathcal{F}) \stackrel{\eqref{random strategy}}{=} H(W_{\mathcal{P}}|A_{1:N}^{[\mathcal{P}]},\mathcal{F})= 0, \quad \forall \mathcal{P} \label{MM-SPIR Reliability}
\end{align}

In order to protect the user's privacy, the query generated to retrieve the set of messages $W_{\mathcal{P}_1}$ should be statistically indistinguishable from the one generated to retrieve the set of messages $W_{\mathcal{P}_2}$ where $|\mathcal{P}_1| = |\mathcal{P}_2| = P$, i.e.,
\begin{align}
\text{[user privacy]} \quad (Q_n^{[\mathcal{P}_1]},A_n^{[\mathcal{P}_1]},W_{1:K},S)  
\sim (Q_n^{[\mathcal{P}_2]},A_n^{[\mathcal{P}_2]},W_{1:K},S), ~
\forall n, ~ \forall \mathcal{P}_1,\mathcal{P}_2 ~ \text{s.t.} ~ |\cp_i|=P   
\label{user-privacy}
\end{align}
The user privacy constraint in \eqref{user-privacy} is equivalent to,
\begin{align}
\text{[user privacy]} \quad I(\mathcal{P};Q_n^{[\mathcal{P}]},A_n^{[\mathcal{P}]},W_{1:K},S) = 0, \quad \forall \cp  
\label{user-privacy-equiv}
\end{align}

In order to protect the databases' privacy, the user should learn nothing about $W_{\bar{\mathcal{P}}}$ which is the complement of $W_{\mathcal{P}}$, i.e., $W_{\bar{\cp}} = W_{1:K} \backslash W_{\cp}$, 
\begin{align}
\text{[database privacy]} \quad I(W_{\bar{\mathcal{P}}};Q_{1:N}^{[\mathcal{P}]},A_{1:N}^{[\mathcal{P}]},\mathcal{F}) = 0,
\quad \forall \mathcal{P} \label{db-privacy}
\end{align}

An achievable MM-SPIR scheme is a scheme that satisfies the MM-SPIR reliability constraint \eqref{MM-SPIR Reliability}, the user privacy constraint \eqref{user-privacy}-\eqref{user-privacy-equiv}, and the database privacy constraint \eqref{db-privacy}. The efficiency of the scheme is measured in terms of the maximal number of downloaded bits by the user from all the databases, denoted by $D_{MM-SPIR}$. Thus, the sum retrieval rate of MM-SPIR is given by
\begin{align} 
R_{MM-SPIR} = \frac{PL}{D_{MM-SPIR}} \label{ratedefinition}  
\end{align}
The sum capacity of MM-SPIR, $C_{MM-SPIR}$, is the supremum of the sum retrieval rates $R_{MM-SPIR}$ over all achievable schemes.

\subsection{MM-SPIR: Main Results} \label{Results}
Our stand-alone result for MM-SPIR is stated in the following theorem. We only consider $N \geq 2$ as SPIR is infeasible for $N=1$.

\begin{theorem}\label{Thm2}
The MM-SPIR capacity for $N\geq 2$, $K \geq 2$, and  a fixed $P \leq K$,  is given by,
\begin{align}
C_{MM-SPIR} = 
 \begin{cases}
 1, & P=K\\
 1 - \frac{1}{N}, & 1 \leq P \leq K-1, ~ H(S) \geq \frac{PL}{N-1} \\
 0, & \text{otherwise}
 \end{cases}
\end{align}
\end{theorem}

The converse proof is given is Section~\ref{converse}, and the achievability proof is given in Section~\ref{achievability}. We have the following remarks concerning Theorem~\ref{Thm2}.

\begin{remark}
The result implies that the capacity of MM-SPIR is exactly the same as the capacity of SM-SPIR \cite{SPIR}. Hence, there is no gain from joint retrieval in comparison to successive single-message SPIR \cite{SPIR}. This in contrast to the gain in MM-PIR \cite{MM-PIR} in comparison to successive single-message PIR \cite{PIR}. MM-SPIR capacity expression in Theorem~\ref{Thm2} inherits all of the structural remarks from \cite{SPIR}.  
\end{remark}

\begin{remark}
Similar to the SM-SPIR problem, we observe a threshold effect on the size of the required common randomness. Specifically, we note that there is a minimal required size for the common randomness above which the problem is feasible. This threshold is $P$ times the threshold in SM-SPIR. Using a common randomness in the amount of the threshold achieves the full capacity, and there is no need to use any more randomness than the threshold.
\end{remark}

\begin{remark}
For the extreme case of $P=K$, the SPIR capacity is $1$ without using any common randomness. This is due to the fact that the user privacy and the database privacy constraints are trivially satisfied, and hence the user can simply download all of the messages from one of the databases without using any common randomness.
\end{remark}

\subsection{MM-SPIR: Converse Proof}\label{converse}
In this section, we derive the converse for Theorem~\ref{Thm2}. In the converse proof, we focus on the case $P\leq K-1$. Because when $P = K$, the trivial upper bound for the retrieval rate $R \leq 1$ and the trivial lower bound for the common randomness $H(S) \geq 0$ suffice. Further, we exclusively focus on the case $K \geq 3$. When $K=1$, we have $P=1$, and the converse trivially follows since $P=K$. When $K=2$: If $P=2$, the converse trivially follows from the converse of $P=K$, and when $P=1$, the converse follows from the converse of SM-SPIR \cite{SPIR}.

Now, focusing on the case $K \geq 3$, and $P\leq K-1$, the total number of possible choices for the index set $\mathcal{P}$ is $\beta = \binom{K}{P}\geq 3$. Thus, there always exist at least three non-identical index sets $\mathcal{P}_1, \mathcal{P}_2, \mathcal{P}_3$ such that $|\mathcal{P}_i|=P$, $i=1,2,3$.

To prove the converse of Theorem~\ref{Thm2}, we first need the following lemmas. Lemmas~\ref{lemma4} are \ref{lemma5} are direct extensions to \cite[Lemmas~1 and 2]{SPIR} to the setting of MM-SPIR. Lemma~\ref{lemma4} simply states that an answer string $A_n^{[\mathcal{P}_1]}$ which is received at the user to retrieve $W_{\mathcal{P}_1}$ has the same size as $A_n^{[\mathcal{P}_2]}$, i.e., all answer strings are symmetric in length, even if we condition over the desired message set $W_{\mathcal{P}_1}$. This lemma is a direct consequence of the user privacy constraint.   

\begin{lemma}[Symmetry] \label{lemma4}
\begin{align}
H(A_n^{[\mathcal{P}_1]}|W_{\mathcal{P}_1},Q_n^{[\mathcal{P}_1]}) &= H(A_n^{[\mathcal{P}_2]}|W_{\mathcal{P}_1},Q_n^{[\mathcal{P}_2]}), \quad  \forall n, ~ \forall \mathcal{P}_1,\mathcal{P}_2 ~ \text{s.t.} ~ \mathcal{P}_1 \neq \mathcal{P}_2, |\mathcal{P}_1| = |\mathcal{P}_2| \label{eq4.1}\\
H(A_n^{[\mathcal{P}_1]}|Q_n^{[\mathcal{P}_1]}) &= H(A_n^{[\mathcal{P}_2]}|Q_n^{[\mathcal{P}_2]}), ~ \quad \quad \quad  \forall n, ~ \forall \mathcal{P}_1,\mathcal{P}_2 ~ \text{s.t.} ~ \mathcal{P}_1 \neq \mathcal{P}_2, |\mathcal{P}_1| = |\mathcal{P}_2| \label{eq4.2}
\end{align}
\end{lemma}

\begin{Proof}
From the user privacy constraint \eqref{user-privacy}, we have
\begin{align}
H(A_n^{[\mathcal{P}_1]},W_{\mathcal{P}_1},Q_n^{[\mathcal{P}_1]}) &= H(A_n^{[\mathcal{P}_2]},W_{\mathcal{P}_1},Q_n^{[\mathcal{P}_2]})\\ H(W_{\mathcal{P}_1},Q_n^{[\mathcal{P}_1]}) &= H(W_{\mathcal{P}_1},Q_n^{[\mathcal{P}_2]})
\end{align}
Using the definition of conditional entropy $H(X|Y)=H(X,Y)-H(Y)$, we obtain \eqref{eq4.1}. The proof of \eqref{eq4.2} follows from the user privacy constraint as well with noting that $H(A_n^{[\mathcal{P}_1]},Q_n^{[\mathcal{P}_1]}) = H(A_n^{[\mathcal{P}_2]},Q_n^{[\mathcal{P}_2]})$ and $ H(A_n^{[\mathcal{P}_1]}) = H(A_n^{[\mathcal{P}_2]})$. 
\end{Proof}

Next, Lemma~\ref{lemma5} states that knowing the user's private randomness $\mathcal{F}$ does not help in decreasing the uncertainty of the answer string $A_n^{[\mathcal{P}]}$.

\begin{lemma}[Effect of conditioning on user's randomness] \label{lemma5}
\begin{align}
H(A_n^{[\mathcal{P}]}|W_{\mathcal{P}},\mathcal{F},Q_n^{[\mathcal{P}]}) = H(A_n^{[\mathcal{P}]}|W_{\mathcal{P}},Q_n^{[\mathcal{P}]}), \quad \forall n, \forall \mathcal{P}
\end{align} 
\end{lemma}

\begin{Proof}
We start with the following mutual information,
\begin{align}
 I(A_n^{[\mathcal{P}]};\mathcal{F}|W_{\mathcal{P}},Q_n^{[\mathcal{P}]})
 &\leq I(A_n^{[\mathcal{P}]},W_{1:K},S;\mathcal{F}|W_{\mathcal{P}},Q_n^{[\mathcal{P}]})\\
 &= I(W_{1:K},S;\mathcal{F}|W_{\mathcal{P}},Q_n^{[\mathcal{P}]}) + I(A_n^{[\mathcal{P}]};\mathcal{F}|W_{1:K},S,W_{\mathcal{P}},Q_n^{[\mathcal{P}]})\\
 &= I(W_{1:K},S;\mathcal{F}|W_{\mathcal{P}},Q_n^{[\mathcal{P}]}) + I(A_n^{[\mathcal{P}]};\mathcal{F}|W_{1:K},S,Q_n^{[\mathcal{P}]})\\
 &= I(W_{1:K},S;\mathcal{F}|W_{\mathcal{P}},Q_n^{[\mathcal{P}]}) + H(A_n^{[\mathcal{P}]}|W_{1:K},S,Q_n^{[\mathcal{P}]})\notag\\
 &\quad- H(A_n^{[\mathcal{P}]}|\mathcal{F},W_{1:K},S,Q_n^{[\mathcal{P}]})\\
 &= I(W_{1:K},S;\mathcal{F}|W_{\mathcal{P}},Q_n^{[\mathcal{P}]}) \label{eq5.1}\\
 &\leq I(W_{1:K},S;\mathcal{F}|W_{\mathcal{P}},Q_n^{[\mathcal{P}]}) + I(W_{\mathcal{P}};\mathcal{F}|Q_n^{[\mathcal{P}]}) \\
 &= I(W_{1:K},W_{\mathcal{P}},S;\mathcal{F}|Q_n^{[\mathcal{P}]}) \\
 &= I(W_{1:K},S;\mathcal{F}|Q_n^{[\mathcal{P}]}) \\
 &\leq I(W_{1:K},S;\mathcal{F}|Q_n^{[\mathcal{P}]}) + I(W_{1:K},S;Q_n^{[\mathcal{P}]}) \\
 &= I(W_{1:K},S;\mathcal{F},Q_n^{[\mathcal{P}]}) \\
 &= 0 \label{eq5.2}
\end{align}
where \eqref{eq5.1} follows from the fact that the answer strings are deterministic functions of the queries and the messages, and \eqref{eq5.2} follows from the independence of $(W_{1:K},S,\mathcal{F})$ and \eqref{random strategy}. Since mutual information cannot be negative, it must be equal to zero, and 
\begin{align}
 H(A_n^{[\mathcal{P}]}|W_{\mathcal{P}},Q_n^{[\mathcal{P}]}) - H(A_n^{[\mathcal{P}]}|W_{\mathcal{P}},\mathcal{F},Q_n^{[\mathcal{P}]}) = I(A_n^{[\mathcal{P}]};\mathcal{F}|W_{\mathcal{P}},Q_n^{[\mathcal{P}]}) = 0
\end{align}
completing the proof.
\end{Proof}

\begin{figure}[t]
	\centering
	\includegraphics[width=0.7\textwidth]{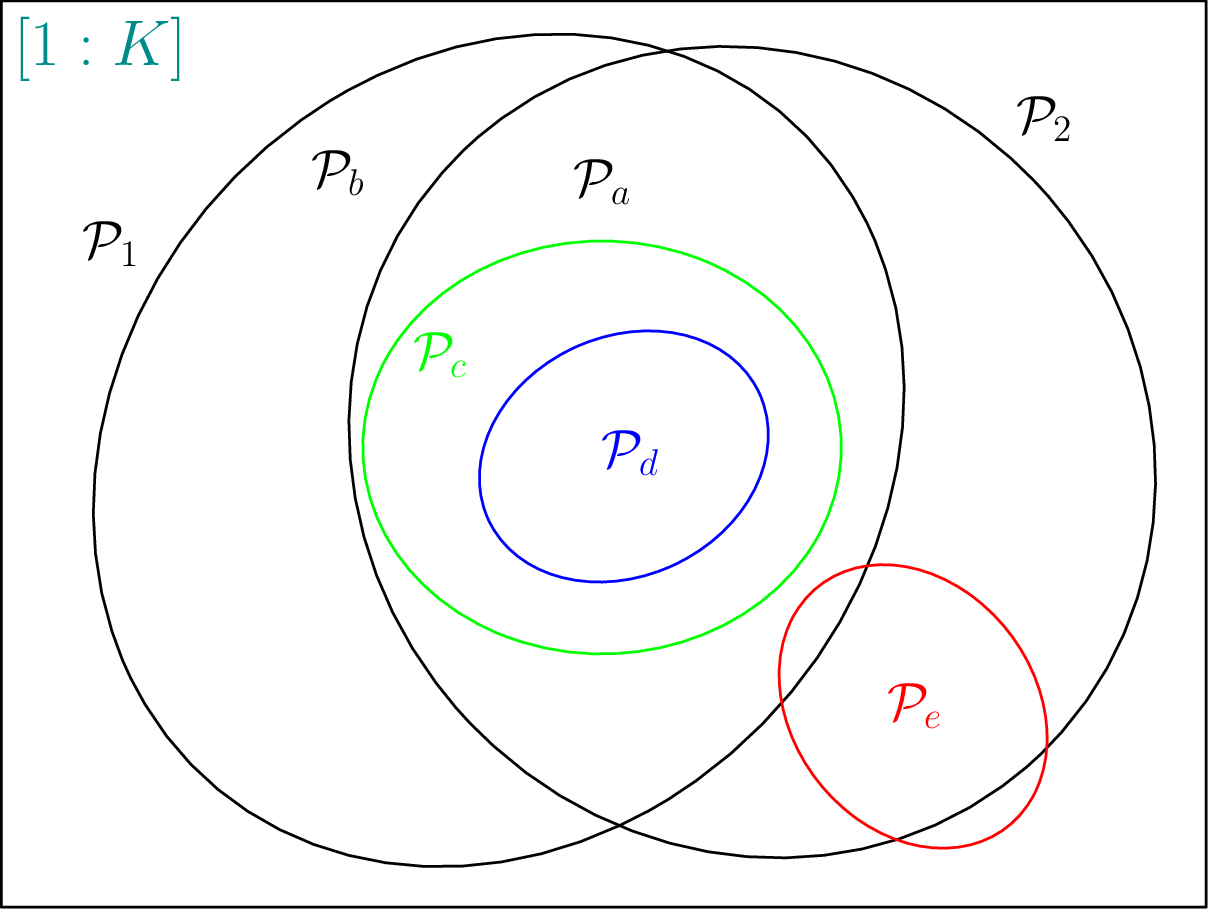}
	\caption{The relation of the index sets presented in Lemma~\ref{lemma-indexsets} and used in Lemmas~\ref{lemma7} and \ref{lemma6}.}
	\label{relation}
\end{figure}

Next, we need Lemma~\ref{lemma-indexsets}, which is an existence proof for index sets with specific properties. This technical lemma is needed in the proofs of upcoming two lemmas, Lemma~\ref{lemma7} and Lemma~\ref{lemma6}. First, we give the definitions of relevant index sets $\mathcal{P}_a$, $\mathcal{P}_b$, $\mathcal{P}_c$, $\mathcal{P}_d$, and an element $i_m$. Given $\mathcal{P}_1$ and $\mathcal{P}_2$, we divide $\mathcal{P}_1$ into two disjoint partitions $\mathcal{P}_a$ and $\mathcal{P}_b$ (i.e., $\mathcal{P}_a \cup \mathcal{P}_b = \mathcal{P}_1$ and $\mathcal{P}_a \cap \mathcal{P}_b = \emptyset$), where $\mathcal{P}_a \subseteq \mathcal{P}_2$ (i.e., $\mathcal{P}_1 \cap \mathcal{P}_2 = \mathcal{P}_a$), $\mathcal{P}_b \subseteq \bar{\mathcal{P}}_2$. Suppose $|\mathcal{P}_a| = M \in [1:P-1]$. Note that since $\cp_1\neq \cp_2$, we cannot have $M=P$. We assume that $\mathcal{P}_a = \{i_1,\cdots,i_M\}$ for clarity of presentation. Given an arbitrary number $m \in [1:M]$, we define a new index set $\mathcal{P}_c = \{i_1,\cdots, i_m\}$ which consists of exactly the first $m$ elements in the index set $\mathcal{P}_a$. Let $i_m$ be the last element from the index set $\mathcal{P}_c$. We obtain a new index set $\mathcal{P}_d = \{i_1,\cdots, i_{m-1}\}$ after removing this element. That means $\mathcal{P}_c = \mathcal{P}_d \cup \{i_m\}$. The relation of all these mentioned index sets is shown in Fig.~\ref{relation}. 

\begin{lemma} \label{lemma-indexsets}
For $K \geq 3$, $1 \leq P \leq K-1$, given index sets $\mathcal{P}_1$, $\mathcal{P}_2$ such that $|\mathcal{P}_i|=P$ for $i=1,2$ and $\mathcal{P}_1 \neq \mathcal{P}_2$, we can construct an index set $\mathcal{P}_3$ such that,
\begin{itemize}
    \item[i)] $\mathcal{P}_3 \neq \mathcal{P}_1$ and $\mathcal{P}_3 \neq \mathcal{P}_2$,
    \item[ii)] $|\mathcal{P}_3|=P$, and
    \item[iii)] $\mathcal{P}_3$ includes $\mathcal{P}_b \cup \mathcal{P}_d$ but does not include the common element $i_m$ in $\mathcal{P}_1 \cap \mathcal{P}_2$.
\end{itemize}
\end{lemma}

\begin{Proof}
The key is to construct an index set $\mathcal{P}_e$ which satisfies the following two constraints: $\mathcal{P}_e \subseteq [1:K] \backslash \{\mathcal{P}_b,\mathcal{P}_c\}$ and $|\mathcal{P}_e| = M-(m-1)$. As we can see, $|\mathcal{P}_a \backslash \mathcal{P}_c| = M-m$ and $|\mathcal{P}_2 \backslash \mathcal{P}_a| \geq 1$. One way to construct the index set $\mathcal{P}_e$ is to include all the $(M-m)$ elements from  the index set $\mathcal{P}_a \backslash \mathcal{P}_c$ and one more element from the index set $\mathcal{P}_2 \backslash \mathcal{P}_a$, i.e.,
\begin{align}
\mathcal{P}_e=(\mathcal{P}_a \backslash \mathcal{P}_c) \cup \{i_*\}  
\end{align}
where $i_* \in \mathcal{P}_2 \backslash \mathcal{P}_a$. The index set $\mathcal{P}_e$ is generally not unique (for some examples, see Examples~\ref{ex1} and \ref{ex2} below). Now, we are ready to construct the index set $\mathcal{P}_3$ as,
\begin{align}
\mathcal{P}_3 = \mathcal{P}_b \cup \mathcal{P}_d \cup \mathcal{P}_e
\end{align}
Since $\mathcal{P}_b$, $\mathcal{P}_d$, $\mathcal{P}_e$ are disjoint sets, $|\mathcal{P}_3|=|\mathcal{P}_b| + |\mathcal{P}_d| + |\mathcal{P}_e| = (P-M) + (m-1) + (M-m+1) = P$. Thus, we are able to construct $\mathcal{P}_3$ such that $|\mathcal{P}_3| = P$. Based on the formulation of $\mathcal{P}_b$, $\mathcal{P}_d$ and $\mathcal{P}_e$, these three index sets do not include the element $i_m$. Hence, $i_m \notin \mathcal{P}_3$. Since both $\mathcal{P}_1$ and $\mathcal{P}_2$ have the element $i_m$ as $i_m$ belongs to their intersection $\mathcal{P}_a$, $\mathcal{P}_3$ is not the same as $\mathcal{P}_1$ or $\mathcal{P}_2$, i.e., $\cp_3\neq \cp_1$, $\cp_3\neq \cp_2$ and $|\cp_3|=P$.
\end{Proof}

The following two examples illustrate the relations between the aforementioned sets, which will be important for the converse proof through the proofs of Lemmas~\ref{lemma7} and \ref{lemma6}.

\begin{example} \label{ex1}
Suppose $K = 3$, $P = 2$ and $N \geq 2$ is an arbitrary positive integer. The total possible number of index sets is $\binom{K}{P} = 3$. Assume $\mathcal{P}_1 = \{1,2\}$, $\mathcal{P}_2 = \{1,3\}$ without loss of generality. Then, $\mathcal{P}_a = \{1\}$, $\mathcal{P}_b = \{2\}$ and the corresponding $M$ is 1. Thus, $m$ can only take the value $1$. That means $\mathcal{P}_c = \{1\}$ and $\mathcal{P}_d$ has to be an empty set. For $\mathcal{P}_e$, we cannot take any element from the set $\mathcal{P}_a \backslash \mathcal{P}_c$ as it is empty, instead we can take the element $3$ from the set $\mathcal{P}_2 \backslash \mathcal{P}_a$. Thus, $\mathcal{P}_e$ is formed as $\{3\}$, and we construct $\mathcal{P}_3 = \{2,3\}$.
\end{example}

\begin{example} \label{ex2}
Suppose $K = 6$, $P = 4$ and $N \geq 2$ is an arbitrary positive integer. The total possible number of index sets is $\binom{K}{P} = 15$. Assume $\mathcal{P}_1 = \{1,3,5,6\}$, $\mathcal{P}_2 = \{2,3,5,6\}$ without loss of generality. Then, $\mathcal{P}_a = \{3,5,6\}$, $\mathcal{P}_b = \{1\}$ and the corresponding $M$ is 3. Thus, $m$ can take the values $1$, $2$ or $3$. To avoid being repetitive, we only consider the cases of $m = 2$ or $m = 3$, which are different from Example~\ref{ex1}. 

When $m = 2$, $\mathcal{P}_c = \{3,5\}$ and $\mathcal{P}_d = \{3\}$. For $\mathcal{P}_e$, we can take the element $6$ from the set $\mathcal{P}_a \backslash \mathcal{P}_c$ and then take the element $2$ from the set $\mathcal{P}_2 \backslash \mathcal{P}_a$. Alternatively, we can pick the element $4$ outside the union $\mathcal{P}_1 \cup \mathcal{P}_2$ instead of the element $6$ from the set $\mathcal{P}_a \backslash \mathcal{P}_c$. Thus, $\mathcal{P}_e$ is formed as $\{2,6\}$ (or $\{4,6\}$). Therefore, we finally obtain $\mathcal{P}_3 = \{1,2,3,6\}$ (or $\{1,3,4,6\}$).

When $m = 3$, $\mathcal{P}_c = \{3,5,6\}$ and $\mathcal{P}_d = \{3,5\}$. For $\mathcal{P}_e$, we cannot take any element from the set $\mathcal{P}_a \backslash \mathcal{P}_c$ since it is empty. We take the element $2$ from the set $\mathcal{P}_2 \backslash \mathcal{P}_a$ or take the element $4$ outside the union $\mathcal{P}_1 \cup \mathcal{P}_2$. Thus, $\mathcal{P}_e$ is formed as $\{2\}$ (or $\{4\}$), and we construct $\mathcal{P}_3 = \{1,2,3,5\}$ (or $\{1,3,4,5\}$).
\end{example}

Next, we need the following lemma. Lemma~\ref{lemma7} states that revealing any individual answer given the messages $(W_{\mathcal{P}_b}, W_{\mathcal{P}_d})$ does not leak any information about the message $W_{i_m}$.

\begin{lemma} [Message leakage within any individual answer string] \label{lemma7}
When $1 \leq P \leq K-1$ and $M \geq 1$, for arbitrary $m \in [1:M]$, the following equality is always true, 
\begin{align} 
H(W_{i_m}|W_{\mathcal{P}_b},W_{\mathcal{P}_d},A_n^{[\mathcal{P}_2]},Q_n^{[\mathcal{P}_2]})
= H(W_{i_m}|W_{\mathcal{P}_b},W_{\mathcal{P}_d},Q_n^{[\mathcal{P}_2]})
\end{align}
\end{lemma}

\begin{remark}
The goal of Lemma~\ref{lemma7} is to prove a key step, equation \eqref{eq6.4}, in the proof of Lemma~\ref{lemma6}. We remark that Lemma~\ref{lemma7} is true for any $m \in [2:M]$ when $M \geq 1$ as proved below. In the case when $m = 1$, the messages set $W_{i_1:i_{m-1}}$ (i.e., $\mathcal{P}_d$) is an empty set and thus Lemma~\ref{lemma7} is still true in this case.
\end{remark}

\begin{Proof}
From the user privacy constraint \eqref{user-privacy}, we have,
\begin{align}
H(W_{\mathcal{P}_b},W_{\mathcal{P}_c},A_n^{[\mathcal{P}_2]},Q_n^{[\mathcal{P}_2]}) 
&= H(W_{\mathcal{P}_b},W_{\mathcal{P}_c},A_n^{[\mathcal{P}_3]},Q_n^{[\mathcal{P}_3]})\\
H(W_{\mathcal{P}_b},W_{\mathcal{P}_d},A_n^{[\mathcal{P}_2]},Q_n^{[\mathcal{P}_2]}) 
&= H(W_{\mathcal{P}_b},W_{\mathcal{P}_d},A_n^{[\mathcal{P}_3]},Q_n^{[\mathcal{P}_3]})
\end{align}
Since $\mathcal{P}_c = \mathcal{P}_d \cup i_m$, we have
\begin{align}
H(W_{i_m}|W_{\mathcal{P}_b},W_{\mathcal{P}_d},A_n^{[\mathcal{P}_2]},Q_n^{[\mathcal{P}_2]}) 
= H(W_{i_m}|W_{\mathcal{P}_b},W_{\mathcal{P}_d},A_n^{[\mathcal{P}_3]},Q_n^{[\mathcal{P}_3]}) \label{eq7.1}
\end{align}
Similarly,
\begin{align}
H(W_{i_m}|W_{\mathcal{P}_b},W_{\mathcal{P}_d},Q_n^{[\mathcal{P}_2]}) 
= H(W_{i_m}|W_{\mathcal{P}_b},W_{\mathcal{P}_d},Q_n^{[\mathcal{P}_3]}) \label{eq7.2}
\end{align}

From the database privacy constraint \eqref{db-privacy}, we have,
\begin{align}
 0 &= I(W_{\bar{\mathcal{P}}_3};A_{1:N}^{[\mathcal{P}_3]},Q_{1:N}^{[\mathcal{P}_3]},\mathcal{F})\\
   &= I(W_{\bar{\mathcal{P}}_3};A_{1:N}^{[\mathcal{P}_3]},W_{\mathcal{P}_3},Q_{1:N}^{[\mathcal{P}_3]},\mathcal{F}) \label{eq7.4}\\
   &\geq I(W_{\bar{\mathcal{P}}_3};A_{1:N}^{[\mathcal{P}_3]},W_{\mathcal{P}_b},W_{\mathcal{P}_d},Q_{1:N}^{[\mathcal{P}_3]}) \label{eq7.5}\\
   &\geq I(W_{i_m};A_{1:N}^{[\mathcal{P}_3]},W_{\mathcal{P}_b},W_{\mathcal{P}_d},Q_{1:N}^{[\mathcal{P}_3]} \label{eq7.6})\\
   &\geq I(W_{i_m};A_n^{[\mathcal{P}_3]},W_{\mathcal{P}_b},W_{\mathcal{P}_d},Q_n^{[\mathcal{P}_3]})\\
   &= I(W_{i_m};A_n^{[\mathcal{P}_3]}|W_{\mathcal{P}_b},W_{\mathcal{P}_d},Q_n^{[\mathcal{P}_3]})\\
   &= H(W_{i_m}|W_{\mathcal{P}_b},W_{\mathcal{P}_d},Q_n^{[\mathcal{P}_3]})
   - H(W_{i_m}|A_n^{[\mathcal{P}_3]},W_{\mathcal{P}_b},W_{\mathcal{P}_d},Q_n^{[\mathcal{P}_3]}) \label{eq7.3}
\end{align}
where \eqref{eq7.4} comes from the MM-SPIR reliability constraint \eqref{MM-SPIR Reliability}, \eqref{eq7.5} comes from the relationship $\mathcal{P}_3 = \mathcal{P}_b \cup \mathcal{P}_d \cup \mathcal{P}_e$ (i.e, $\mathcal{P}_b \cup \mathcal{P}_d \subseteq \mathcal{P}_3$), and \eqref{eq7.6} comes from the relationship $i_m \in \bar{\mathcal{P}}_3$. Thus, $H(W_{i_m}|W_{\mathcal{P}_b},W_{\mathcal{P}_d},Q_n^{[\mathcal{P}_3]}) \leq  H(W_{i_m}|A_n^{[\mathcal{P}_3]},W_{\mathcal{P}_b},W_{\mathcal{P}_d},Q_n^{[\mathcal{P}_3]})$. This concludes the proof by observing that $H(W_{i_m}|W_{\mathcal{P}_b},W_{\mathcal{P}_d},Q_n^{[\mathcal{P}_3]}) \geq  H(W_{i_m}|A_n^{[\mathcal{P}_3]},W_{\mathcal{P}_b},W_{\mathcal{P}_d},Q_n^{[\mathcal{P}_3]})$ trivially as conditioning cannot increase entropy.
\end{Proof}

Finally, the following lemma states that conditioning on an undesired message set does not decrease the uncertainty on any individual answer string. This is a consequence of the database privacy constraint.

\begin{lemma}[Effect of conditioning on an undesired message set] \label{lemma6}
\begin{align}
H(A_n^{[\mathcal{P}_2]}|W_{\mathcal{P}_1},Q_n^{[\mathcal{P}_2]}) = H(A_n^{[\mathcal{P}_2]}|Q_n^{[\mathcal{P}_2]}), \quad \forall n, ~ \forall \mathcal{P}_1,\mathcal{P}_2 ~ \text{s.t.} ~ \mathcal{P}_1 \neq \mathcal{P}_2, |\mathcal{P}_1| = |\mathcal{P}_2| \label{lemma6-eqn}
\end{align}
\end{lemma}

\begin{remark}
We note that although Lemma~\ref{lemma6} has the same flavor as \cite[eqn.~(39)]{SPIR}, the proof is much more involved. The main reason for this difficulty is the inter-relations between subsets of messages of size $P$. Specifically, in SM-SPIR, all message subsets are of size $P=1$, and therefore, they are disjoint. However, in MM-SPIR, the message subsets are of size $P$, and they intersect in general, i.e., for a given $\mathcal{P}_1$, $\mathcal{P}_2$ such that $|\mathcal{P}_1|=|\mathcal{P}_2|=P$, the intersection $\mathcal{P}_1 \cap \mathcal{P}_2$ is not an empty set in general in contrast to SM-SPIR. Dealing with message subset intersections is the essence of introducing and proving Lemmas~\ref{lemma-indexsets}, \ref{lemma7} and \ref{lemma6}. 
\end{remark}

\begin{Proof}
From the database privacy constraint \eqref{db-privacy}, we have,
\begin{align}
 0 &= I(W_{\bar{\mathcal{P}}_2};A_{1:N}^{[\mathcal{P}_2]},Q_{1:N}^{[\mathcal{P}_2]},\mathcal{F})\\
   &\geq I(W_{\bar{\mathcal{P}}_2};A_n^{[\mathcal{P}_2]},Q_n^{[\mathcal{P}_2]})\\
   &\geq I(W_{\mathcal{P}_b};A_n^{[\mathcal{P}_2]},Q_n^{[\mathcal{P}_2]}) \label{eq6.1}\\
   &= I(W_{\mathcal{P}_b};A_n^{[\mathcal{P}_2]}|Q_n^{[\mathcal{P}_2]})\label{eq6.11}\\
   &= H(W_{\mathcal{P}_b}|Q_n^{[\mathcal{P}_2]}) - H(W_{\mathcal{P}_b}|A_n^{[\mathcal{P}_2]},Q_n^{[\mathcal{P}_2]}) \label{eq6.2}
\end{align}
where \eqref{eq6.1} comes from the relationship $\mathcal{P}_b \subseteq \bar{\mathcal{P}}_2$, \eqref{eq6.11} follows from the independence of messages and queries. Hence, $H(W_{\mathcal{P}_b}|Q_n^{[\mathcal{P}_2]})= H(W_{\mathcal{P}_b}|A_n^{[\mathcal{P}_2]},Q_n^{[\mathcal{P}_2]})$ as the reverse implication follows form the fact that conditioning cannot increase entropy.

\paragraph{Case~1: $M=0$:}
In this case, there is no intersection between $\mathcal{P}_1$ and $\mathcal{P}_2$. $W_{\mathcal{P}_a}$ is an empty set of messages and then $W_{\mathcal{P}_1} = W_{\mathcal{P}_b}$. Hence,
\begin{align}
  I(W_{\mathcal{P}_1};A_n^{[\mathcal{P}_2]}|Q_n^{[\mathcal{P}_2]})=I(W_{\mathcal{P}_b};A_n^{[\mathcal{P}_2]}|Q_n^{[\mathcal{P}_2]})=0  \label{eq6.7} 
\end{align}
where \eqref{eq6.7} follows from \eqref{eq6.11}. This proves (\ref{lemma6-eqn}), the claim of lemma, when $M=0$.

\paragraph{Case~2: $M \geq 1$:}
In this case, $W_{\mathcal{P}_1} = W_{\mathcal{P}_a} \cup W_{\mathcal{P}_b}$ and $W_{\mathcal{P}_a} = \{W_{i_1},\cdots,W_{i_M}\}$.
\begin{align}
H(W_{\mathcal{P}_a}|W_{\mathcal{P}_b},A_n^{[\mathcal{P}_2]},Q_n^{[\mathcal{P}_2]})
&= H(W_{i_1:i_M}|W_{\mathcal{P}_b},A_n^{[\mathcal{P}_2]},Q_n^{[\mathcal{P}_2]})\\
&= H(W_{i_1}|W_{\mathcal{P}_b},A_n^{[\mathcal{P}_2]},Q_n^{[\mathcal{P}_2]})
+ H(W_{i_2}|W_{i_1},W_{\mathcal{P}_b},A_n^{[\mathcal{P}_2]},Q_n^{[\mathcal{P}_2]}) \notag\\ 
&\qquad + \cdots
+ H(W_{i_M}|W_{i_1:i_M-1},W_{\mathcal{P}_b},A_n^{[\mathcal{P}_2]},Q_n^{[\mathcal{P}_2]})\\
&= H(W_{i_1}|W_{\mathcal{P}_b},Q_n^{[\mathcal{P}_2]})
+ H(W_{i_2}|W_{i_1},W_{\mathcal{P}_b},Q_n^{[\mathcal{P}_2]}) \notag\\ 
&\qquad + \cdots
+ H(W_{i_M}|W_{i_1:i_M-1},W_{\mathcal{P}_b},Q_n^{[\mathcal{P}_2]}) \hspace{3mm} \label{eq6.4}\\
&= H(W_{i_1:i_M}|W_{\mathcal{P}_b},Q_n^{[\mathcal{P}_2]})\\
&= H(W_{\mathcal{P}_a}|W_{\mathcal{P}_b},Q_n^{[\mathcal{P}_2]}) \label{eq6.5}
\end{align}
where \eqref{eq6.4} comes from the direct application of Lemma \ref{lemma7}.

Thus, we have,
\begin{align}
I(W_{\mathcal{P}_1};A_n^{[\mathcal{P}_2]}|Q_n^{[\mathcal{P}_2]}) 
&= H(W_{\mathcal{P}_1}|Q_n^{[\mathcal{P}_2]}) - H(W_{\mathcal{P}_1}|A_n^{[\mathcal{P}_2]},Q_n^{[\mathcal{P}_2]})\\
&= H(W_{\mathcal{P}_1}|Q_n^{[\mathcal{P}_2]}) - H(W_{\mathcal{P}_a},W_{\mathcal{P}_b}|A_n^{[\mathcal{P}_2]},Q_n^{[\mathcal{P}_2]})\\
&= H(W_{\mathcal{P}_1}|Q_n^{[\mathcal{P}_2]}) - H(W_{\mathcal{P}_b}|A_n^{[\mathcal{P}_2]},Q_n^{[\mathcal{P}_2]}) -  H(W_{\mathcal{P}_a}|W_{\mathcal{P}_b},A_n^{[\mathcal{P}_2]},Q_n^{[\mathcal{P}_2]})\\
&= H(W_{\mathcal{P}_1}|Q_n^{[\mathcal{P}_2]}) - H(W_{\mathcal{P}_b}|Q_n^{[\mathcal{P}_2]}) -  H(W_{\mathcal{P}_a}|W_{\mathcal{P}_b},Q_n^{[\mathcal{P}_2]}) \label{eq6.6}\\
&= H(W_{\mathcal{P}_1}|Q_n^{[\mathcal{P}_2]}) - H(W_{\mathcal{P}_a},W_{\mathcal{P}_b}|Q_n^{[\mathcal{P}_2]})\\
&= H(W_{\mathcal{P}_1}|Q_n^{[\mathcal{P}_2]}) - H(W_{\mathcal{P}_1}|Q_n^{[\mathcal{P}_2]}) \\
&= 0 \label{eq6.8}
\end{align}
where \eqref{eq6.6} follows from \eqref{eq6.2} and \eqref{eq6.5}. This proves (\ref{lemma6-eqn}), the claim of lemma, when $M\geq 1$.

Combining \eqref{eq6.7} and \eqref{eq6.8} proves (\ref{lemma6-eqn}) in all cases completing the proof.
\end{Proof}

\begin{remark}
    The intuition behind Lemma~\ref{lemma6} is as follows: If the pair $(A_n^{[\mathcal{P}_2]},Q_n^{[\mathcal{P}_2]})$ provide any information about $W_{\mathcal{P}_1}$, they have to provide some information about $W_{\bar{\mathcal{P}}_1}$ under the user privacy constraint. However, database privacy constraint is thus obviously violated if the user receives any information about $W_{\bar{\mathcal{P}}_1}$. Consequently, the pair $(A_n^{[\mathcal{P}_2]},Q_n^{[\mathcal{P}_2]})$ can never provide any information about $W_{\bar{\mathcal{P}}_1}$. Therefore, we are able to derive $H(W_{\mathcal{P}_1}|A_n^{[\mathcal{P}_2]},Q_n^{[\mathcal{P}_2]}) = H(W_{\mathcal{P}_1}) \stackrel{\eqref{queries and messages}}{=} H(W_{\mathcal{P}_1}|Q_n^{[\mathcal{P}_2]})$, and hence $I(W_{\mathcal{P}_1};A_n^{[\mathcal{P}_2]}|Q_n^{[\mathcal{P}_2]}) = 0$.
\end{remark}

Now, we are ready to construct the main body of the converse proof for MM-SPIR, as well as the minimal entropy of common randomness required to achieve perfect MM-SPIR. Since we dealt with the inter-relations between message subsets in the previous lemmas and reached similar conclusions to those in SM-SPIR \cite{SPIR}, the main body of the converse proof will be similar in structure to its counterpart in SM-SPIR.

\textbf{The proof for $R \leq C_{MM-SPIR}$:}
\begin{align}
 PL &= H(W_{\mathcal{P}_1})\\
   &= H(W_{\mathcal{P}_1}|\mathcal{F}) \label{th1.9}\\
   &= H(W_{\mathcal{P}_1}|\mathcal{F}) - H(W_{\mathcal{P}_1}|A_{1:N}^{[\mathcal{P}_1]},\mathcal{F}) \label{th1.1}\\
   &= I(W_{\mathcal{P}_1};A_{1:N}^{[\mathcal{P}_1]}|\mathcal{F})\\
   &= H(A_{1:N}^{[\mathcal{P}_1]}|\mathcal{F}) - H(A_{1:N}^{[\mathcal{P}_1]}|W_{\mathcal{P}_1},\mathcal{F})\\
   &= H(A_{1:N}^{[\mathcal{P}_1]}|\mathcal{F}) - H(A_{1:N}^{[\mathcal{P}_1]}|W_{\mathcal{P}_1},\mathcal{F},Q_n^{[\mathcal{P}_1]} \label{th1.2})\\
   &\leq H(A_{1:N}^{[\mathcal{P}_1]}|\mathcal{F}) - H(A_n^{[\mathcal{P}_1]}|W_{\mathcal{P}_1},\mathcal{F},Q_n^{[\mathcal{P}_1]})\\
   &= H(A_{1:N}^{[\mathcal{P}_1]}|\mathcal{F}) - H(A_n^{[\mathcal{P}_1]}|W_{\mathcal{P}_1},Q_n^{[\mathcal{P}_1]}) \label{th1.3} \\
   &= H(A_{1:N}^{[\mathcal{P}_1]}|\mathcal{F}) - H(A_n^{[\mathcal{P}_2]}|W_{\mathcal{P}_1},Q_n^{[\mathcal{P}_2]}) \label{th1.4}\\
   &= H(A_{1:N}^{[\mathcal{P}_1]}|\mathcal{F}) - H(A_n^{[\mathcal{P}_2]}|Q_n^{[\mathcal{P}_2]}) \label{th1.5}\\
   &= H(A_{1:N}^{[\mathcal{P}_1]}|\mathcal{F}) - H(A_n^{[\mathcal{P}_1]}|Q_n^{[\mathcal{P}_1]}) \label{th1.6}\\
   &\leq H(A_{1:N}^{[\mathcal{P}_1]}|\mathcal{F}) - H(A_n^{[\mathcal{P}_1]}|Q_n^{[\mathcal{P}_1]},\mathcal{F})\\
   &= H(A_{1:N}^{[\mathcal{P}_1]}|\mathcal{F}) - H(A_n^{[\mathcal{P}_1]}|\mathcal{F}) \label{th1.7}
\end{align}
where \eqref{th1.9} follows from the independence of the user's private randomness and the messages, \eqref{th1.1} follows from the MM-SPIR reliability constraint \eqref{MM-SPIR Reliability}, \eqref{th1.2} follows from the fact that the queries are deterministic functions of the user's private randomness $\mathcal{F}$ \eqref{random strategy}, \eqref{th1.3} follows from Lemma~\ref{lemma5}, \eqref{th1.4} follows from the first part of Lemma~\ref{lemma4}, \eqref{th1.5} follows from Lemma~\ref{lemma6}, \eqref{th1.6} follows from the second part Lemma~\ref{lemma4}, and \eqref{th1.7} again follows from the fact that the queries are deterministic functions of the user's private randomness $\mathcal{F}$ \eqref{random strategy}.

By summing \eqref{th1.7} up for all $n\in[1:N]$ and letting $\mathcal{P}$ denote the general desired index set, we obtain,
\begin{align}
 NPL &\leq NH(A_{1:N}^{[\mathcal{P}]}|\mathcal{F}) - \sum\limits_{n=1}^N H(A_n^{[\mathcal{P}]}|\mathcal{F})\\
    &\leq NH(A_{1:N}^{[\mathcal{P}]}|\mathcal{F}) - H(A_{1:N}^{[\mathcal{P}]}|\mathcal{F})\\
    &= (N-1)H(A_{1:N}^{[\mathcal{P}]}|\mathcal{F}) \label{th1.8}\\
    &\leq (N-1)\sum_{n=1}^N H(A_n^{[\mathcal{P}]}|\mathcal{F})\\
    &\leq (N-1)\sum_{n=1}^N H(A_n^{[\mathcal{P}]}) \label{before-rateresult}
\end{align}
which leads to the desired converse result on the retrieval rate,
\begin{align}
 R_{MM-SPIR} = \frac{PL}{D_{MM-SPIR} \leq \frac{PL}{\sum_{n=1}^N H(A_n^{[\mathcal{P}]})} \leq \frac{N-1}{N} = 1-\frac{1}{N}} \label{rateresult}  
\end{align}

\textbf{The proof for $H(S) \geq \frac{PL}{N-1}$:} 
\begin{align}
 0 &= I(W_{\bar{\mathcal{P}}_1};A_{1:N}^{[\mathcal{P}_1]},Q_{1:N}^{[\mathcal{P}_1]},\mathcal{F}) \label{cr1.1}\\
   &\geq I(W_{\bar{\mathcal{P}}_1};A_{1:N}^{[\mathcal{P}_1]},\mathcal{F}) \\
   &= I(W_{\bar{\mathcal{P}}_1};A_{1:N}^{[\mathcal{P}_1]},W_{\mathcal{P}_1},\mathcal{F}) \label{cr1.2}\\
   &= I(W_{\bar{\mathcal{P}}_1};A_{1:N}^{[\mathcal{P}_1]}|W_{\mathcal{P}_1},\mathcal{F}) \\
   &\geq I(W_{\bar{\mathcal{P}}_1};A_n^{[\mathcal{P}_1]}|W_{\mathcal{P}_1},\mathcal{F}) \\
   &= H(A_n^{[\mathcal{P}_1]}|W_{\mathcal{P}_1},\mathcal{F}) - H(A_n^{[\mathcal{P}_1]}|W_{1:K},\mathcal{F}) \\
   &= H(A_n^{[\mathcal{P}_1]}|W_{\mathcal{P}_1},\mathcal{F}) - H(A_n^{[\mathcal{P}_1]}|W_{1:K},\mathcal{F}) + H(A_n^{[\mathcal{P}_1]}|W_{1:K},\mathcal{F},S) \label{cr1.3}\\
   &= H(A_n^{[\mathcal{P}_1]}|W_{\mathcal{P}_1},\mathcal{F}) - I(S;A_n^{[\mathcal{P}_1]}|W_{1:K},\mathcal{F}) \\
   &= H(A_n^{[\mathcal{P}_1]}|W_{\mathcal{P}_1},\mathcal{F}) - H(S|W_{1:K},\mathcal{F}) + H(S|A_n^{[\mathcal{P}_1]},W_{1:K},\mathcal{F}) \\
   &= H(A_n^{[\mathcal{P}_1]}|W_{\mathcal{P}_1},\mathcal{F}) - H(S) + H(S|A_n^{[\mathcal{P}_1]},W_{1:K},\mathcal{F}) \label{cr1.4}\\
   &\geq H(A_n^{[\mathcal{P}_1]}|W_{\mathcal{P}_1},\mathcal{F}) - H(S)\\
   &= H(A_n^{[\mathcal{P}_1]}|W_{\mathcal{P}_1},\mathcal{F},Q_n^{[\mathcal{P}_1]}) - H(S) \label{cr1.5}\\
   &= H(A_n^{[\mathcal{P}_1]}|Q_n^{[\mathcal{P}_1]}) - H(S) \label{cr1.6} 
\end{align}
where \eqref{cr1.1} follows from the database privacy constraint \eqref{db-privacy}, \eqref{cr1.2} follows from the MM-SPIR reliability constraint \eqref{MM-SPIR Reliability}, \eqref{cr1.3} follows from the fact that the answer strings are deterministic functions of messages and queries which are also functions of the randomness $\mathcal{F}$ as in \eqref{random strategy} and \eqref{determined answer string}, \eqref{cr1.4} follows from the independence of the common randomness, messages, and user's private randomness as in \eqref{all independent}, \eqref{cr1.5} follows from \eqref{random strategy}, and \eqref{cr1.6} follows from the steps between \eqref{th1.3}-\eqref{th1.6} by applying Lemma~\ref{lemma4}, \ref{lemma5} and \ref{lemma6} again.

By summing \eqref{cr1.6} up for all $n\in[1:N]$ and letting $\mathcal{P}$ denote the general desired index set again, we obtain, 
\begin{align}
 0 &\geq \sum\limits_{n=1}^N H(A_n^{[\mathcal{P}]}|Q_n^{[\mathcal{P}]}) - NH(S)\\
   &\geq H(A_{1:N}^{[\mathcal{P}]}|Q_n^{[\mathcal{P}]}) - NH(S)\\
   &\geq H(A_{1:N}^{[\mathcal{P}]}|Q_n^{[\mathcal{P}]},\mathcal{F}) - NH(S)\\
   &= H(A_{1:N}^{[\mathcal{P}]}|\mathcal{F}) - NH(S) \label{cr1.7}\\
   &\geq \frac{N}{N-1}PL - NH(S) \label{cr1.8}
\end{align}
where \eqref{cr1.7} follows from \eqref{random strategy} and \eqref{cr1.8} follows from \eqref{th1.8}, which leads to a lower bound for the minimal required entropy of common randomness $S$,
\begin{align}
 H(S) \geq \frac{PL}{N-1} \label{crresult}        
\end{align}

\subsection{MM-SPIR: Achievability Proof}\label{achievability}
Since the MM-SPIR capacity is the same as the SM-SPIR capacity, and the required common randomness is $P$ times the required common randomness for SM-SPIR, we can use the achievable scheme in \cite{SPIR} successively $P$ times in a row (by utilizing independent common randomness each time) to achieve the MM-SPIR capacity. Although the query structure for the capacity-achieving scheme for SPIR in \cite{SPIR} is quite simple, it is fundamentally different than the query structure for the capacity-achieving scheme for PIR in \cite{PIR}. This means that user/databases should execute different query structures for different database privacy levels. In this paper, by combining ideas for achievability from \cite{MM-PIR} and \cite{SPIR_Mismatched}, we propose an alternative capacity-achieving scheme for MM-SPIR for any\footnote{We note that the capacity-achieving scheme for $K=P$ is simply to download all messages from one of the databases, hence, without loss of generality, we focus on the case $1 \leq P \leq K-1$ in this section.} $P$. Our achievability scheme enables us to switch between MM-PIR and MM-SPIR seamlessly, and therefore support different database privacy levels, as the basic query structures are similar. We start with two motivating examples in Section~\ref{mot_example}, give the general achievable scheme in Section~\ref{achi_MMSPIR}, and calculate its rate and required common randomness amount in Section~\ref{rate-calculation}. 

For convenience, we use the \emph{$k$-sum} notation in \cite{PIR, MM-PIR}. A $k$-sum is a sum of $k$ symbols from $k$ different messages. Thus, a $k$-sum symbol appears only in round $k$. We denote the number of stages in round $k$ by $\alpha_k$, which was originally introduced in \cite{MM-PIR}. In addition, we use $\nu$ to denote the number of repetitions of the scheme\footnote{\label{footnote6} When we refer to the scheme in \cite{MM-PIR}, we refer to the near-optimal scheme in \cite{MM-PIR} which was introduced for $K/P \geq 2$. Reference \cite{MM-PIR} has a different, optimal, scheme for $K/P\leq 2$. However, in this paper, even when $K/P\leq 2$, we still refer to (and use) the near-optimal scheme in \cite{MM-PIR}.} in \cite{MM-PIR} we need before we start assigning common randomness symbols.

\subsubsection{Motivating Examples} \label{mot_example}
\begin{example} \label{example3}
Consider the case $K = 3$, $P = 1$, $N = 3$. Our achievable scheme is as follows: First, we generate an initial query table, which strictly follows the query table generation in \cite{MM-PIR}. For this case, from \cite{MM-PIR}, we obtain the number of stages needed in each round as $\alpha_1 = 1, \alpha_2 = 2, \alpha_3 = 4$. From the perspective of a database, before the assignment of common randomness symbols begins, the total number of downloaded desired symbols in round $1$ is $\alpha_1 P = 1\times 1= 1$. Thus, we need $1$ previously downloaded common randomness symbol for this desired symbol. Since this common randomness symbol needs to come from the other $N-1=2$ databases, the required common randomness to be downloaded from each database is $\frac{1}{2}$ symbols (assuming a symmetric scheme that distributes downloads equally over the other $2$ databases). Thus, in order to obtain an integer number of common randomness symbols to be downloaded from each database, we repeat the scheme in \cite{MM-PIR} two times (i.e., $\nu = 2$) before we begin assigning the common randomness symbols. Hence, the number of stages in each round become $\nu \alpha_k=2 \alpha_k$, for $k=1, 2, 3$. That is we have $2$ stages of $1$-sums, $4$ stages of $2$-sums and $8$ stages of $3$-sums; see Table~\ref{table6.1}. 

\begin{table}[ht]
\begin{center}
\begin{tabular}{ |c|c|c| }
 \hline
 Database 1 & Database 2 & Database 3 \\
 \hline
 $s_1$ & $s_2$ & $s_3$\\
 \hline
 $a_1+s_2$ & $a_3+s_1$ & $a_5+s_1$ \\
 $a_2+s_3$ & $a_4+s_3$ & $a_6+s_2$ \\ 
 $b_1+s_4$ & $b_3+s_8$ & $b_5+s_{12}$ \\
 $b_2+s_5$ & $b_4+s_9$ & $b_6+s_{13}$ \\
 $c_1+s_6$ & $c_3+s_{10}$ & $c_5+s_{14}$ \\
 $c_2+s_7$ & $c_4+s_{11}$ & $c_6+s_{15}$ \\
 \hline
 $a_7+b_3+s_8$ & $a_{15}+b_1+s_4$ & $a_{23}+b_1+s_4$\\
 $a_8+b_4+s_9$ & $a_{16}+b_2+s_5$ & $a_{24}+b_2+s_5$\\
 $a_9+b_5+s_{12}$ & $a_{17}+b_5+s_{12}$ & $a_{25}+b_3+s_8$\\
 $a_{10}+b_6+s_{13}$ & $a_{18}+b_6+s_{13}$ & $a_{26}+b_4+s_9$\\
 $a_{11}+c_3+s_{10}$ & $a_{19}+c_1+s_6$ & $a_{27}+c_1+s_6$\\
 $a_{12}+c_4+s_{11}$ & $a_{20}+c_2+s_7$ & $a_{28}+c_2+s_7$\\
 $a_{13}+c_5+s_{14}$ & $a_{21}+c_5+s_{14}$ & $a_{29}+c_3+s_{10}$\\
 $a_{14}+c_6+s_{15}$ & $a_{22}+c_6+s_{15}$ & $a_{30}+c_4+s_{11}$\\
 $b_7+c_7+s_{16}$ & $b_{11}+c_{11}+s_{20}$ & $b_{15}+c_{15}+s_{24}$\\
 $b_8+c_8+s_{17}$ & $b_{12}+c_{12}+s_{21}$ & $b_{16}+c_{16}+s_{25}$\\
 $b_9+c_9+s_{18}$ & $b_{13}+c_{13}+s_{22}$ & $b_{17}+c_{17}+s_{26}$\\
 $b_{10}+c_{10}+s_{19}$ & $b_{14}+c_{14}+s_{23}$ & $b_{18}+c_{18}+s_{27}$\\
 \hline
 $a_{31}+b_{11}+c_{11}+s_{20}$ & $a_{39}+b_{7}+c_{7}+s_{16}$ & $a_{47}+b_{7}+c_{7}+s_{16}$\\ 
 $a_{32}+b_{12}+c_{12}+s_{21}$ & $a_{40}+b_{8}+c_{8}+s_{17}$ & $a_{48}+b_{8}+c_{8}+s_{17}$\\
 $a_{33}+b_{13}+c_{13}+s_{22}$ & $a_{41}+b_{9}+c_{9}+s_{18}$ & $a_{49}+b_{9}+c_{9}+s_{18}$\\
 $a_{34}+b_{13}+c_{14}+s_{23}$ & $a_{42}+b_{10}+c_{10}+s_{19}$ & $a_{50}+b_{10}+c_{10}+s_{19}$\\
 $a_{35}+b_{15}+c_{15}+s_{24}$ & $a_{43}+b_{15}+c_{15}+s_{24}$ & $a_{51}+b_{11}+c_{11}+s_{20}$\\ 
 $a_{36}+b_{16}+c_{16}+s_{25}$ & $a_{44}+b_{16}+c_{16}+s_{25}$ & $a_{52}+b_{12}+c_{12}+s_{21}$\\
 $a_{37}+b_{17}+c_{17}+s_{26}$ & $a_{45}+b_{17}+c_{17}+s_{26}$ & $a_{53}+b_{13}+c_{13}+s_{22}$\\
 $a_{38}+b_{18}+c_{18}+s_{27}$ & $a_{46}+b_{18}+c_{18}+s_{27}$ & $a_{54}+b_{14}+c_{14}+s_{23}$\\
 \hline
\end{tabular}
\end{center}
\vspace*{-0.4cm}
\caption{The query table for the case $K = 3, P = 1, N = 3$.}
\label{table6.1}
\end{table}

We are now ready to start assigning the common randomness symbols. We first download $1$ common randomness symbol from each database; for instance, we download $s_1$ from database $1$. In round $1$, we mix (i.e., add) a common randomness symbol to each $1$-sum. All the common randomness symbols at each database should be distinct; for instance, observe that, we add $s_2, s_3, s_4, s_5, s_6, s_7$ at database $1$. Second, the common randomness symbols added to the desired symbols ($a$ symbols in this example) must be downloaded from other databases; for instance, note that $s_2$ and $s_3$ added to symbols $a_1$ and $a_2$ are downloaded from databases $2$ and $3$. Note that the indices of the common randomness symbols added to the undesired symbols (symbols $b$ and $c$) increase cumulatively, e.g., $s_4, s_5, s_6, s_7$ at database $1$ in round $1$, and these symbols are not separately downloaded by the user. 

In round $2$, for every $2$-sum containing a desired message symbol, we add a side information symbol downloaded from another database which already contains a common randomness symbol; for instance, we add $b_3+s_8$ that is already downloaded from database $2$, to the desired symbol $a_7$ at database $1$, i.e., we download $a_7+b_3+s_8$. On the other hand, for every $2$-sum not containing any desired message symbols, we add a new distinct common randomness symbol with a cumulatively increasing index; for instance, for the download $b_7+c_7$ from database $1$, we add $s_{16}$ which is a new non-downloaded common randomness symbol, and download $b_7+c_7+s_{16}$. Finally, in round $3$, where we download $3$-sums, and hence every download contains a desired symbol, we add the side information generated at other databases; for instance, we add $b_{11}+c_{11}+s_{20}$ downloaded from database $2$, to $a_{31}$ and download $a_{31}+b_{11}+c_{11}+s_{20}$. This completes the achievable scheme for this case. The complete query table is shown in Table~\ref{table6.1}.

Now, we calculate the rate of this scheme. The length of each message is $L=54$, and the total number of downloads is $D=81$. Thus, the rate $R$ of this scheme is $\frac{54}{81}=\frac{2}{3}=1-\frac{1}{3}$, which matches the capacity expression. In addition, we used $27$ common randomness symbols, hence the required common randomness $H(S)$ is $27=\frac{54}{2}$, which matches the required minimum.
\end{example}

\begin{example} \label{example4}
Consider the case $K = 5, P = 3, N = 2$. Our achievable scheme is as follows: Again, first, we generate an initial query table, which strictly follows the query table generation in \cite{MM-PIR}. Note that, we still use the near-optimal scheme in \cite{MM-PIR}, even though for this case $K/P\leq 2$ (see Footnote~\ref{footnote6}). For this case, from \cite{MM-PIR}, we obtain the number of stages needed in each round as $\alpha_1 = 3$, $\alpha_2 = 1$, $\alpha_3 = \alpha_4 = 0$ and $\alpha_5 = 1$. In this case, from the perspective of a database, before the assignment of common randomness symbols begins, the total number of downloaded desired symbols in round $1$ is $\alpha_1 P = 3\times 3=9$. Thus, we need $9$ previously downloaded common randomness symbols for these desired symbols. These common randomness symbols need to come from the other $N-1=1$ database. In this case, since $9/1=9$ is an integer already, we do not need to repeat the scheme unlike the case in Example~\ref{example3}. Thus, $\nu=1$ here, there is no need for repetition, and the underlying query structure before adding common randomness symbols is exactly the same as \cite{MM-PIR}; see Table~\ref{table6.2}. 

We are now ready to start assigning the common randomness symbols. We first download $9$ common randomness symbols from each database; for instance, we download $s_1,\cdots,s_9$ from database $1$. In round $1$, we add a common randomness symbol to each $1$-sum. All the common randomness symbols at each database should be distinct; for instance, observe that, we add $s_{10},\cdots,s_{24}$ at database $1$. Second, the common randomness symbols added to the desired symbols ($a$, $b$, $c$ symbols in this example) must be downloaded from the other databases; for instance, note that $s_{10}, \cdots, s_{18}$ added to symbols $a_1, b_1, c_1, a_2, b_2, c_2, a_3, b_3, c_3$ are downloaded from database $2$. Note that the indices of the common randomness symbols added to the undesired symbols (symbols $d$ and $e$) increase cumulatively, e.g., $s_{19} \cdots, s_{24}$ at database $1$ in round $1$, and these symbols are not separately downloaded by the user. 

\begin{table}[ht]
\begin{center}
\begin{tabular}{ |c|c| }
 \hline
 Database 1 & Database 2 \\
 \hline
 $s_1, s_2, s_3$ & $s_{10}, s_{11}, s_{12}$\\
 $s_4, s_5, s_6$ & $s_{13}, s_{14}, s_{15}$\\
 $s_7, s_8, s_9$ & $s_{16}, s_{17}, s_{18}$\\
 \hline
 $s_{31}, s_{32}, s_{33}$ & $s_{34}, s_{35}, s_{36}$\\
 \hline
 $a_1+s_{10}$ & $a_{4}+s_1$\\
 $b_1+s_{11}$ & $b_{4}+s_2$\\
 $c_1+s_{12}$ & $c_{4}+s_3$\\
 $d_1+s_{19}$ & $d_{4}+s_{25}$\\
 $e_1+s_{20}$ & $e_{4}+s_{26}$\\ 
 $a_2+s_{13}$ & $a_{5}+s_4$\\
 $b_2+s_{14}$ & $b_{5}+s_5$\\
 $c_2+s_{15}$ & $c_{5}+s_6$\\
 $d_2+s_{21}$ & $d_{5}+s_{27}$\\
 $e_2+s_{22}$ & $e_{5}+s_{28}$\\
 $a_3+s_{16}$ & $a_{6}+s_7$\\
 $b_3+s_{17}$ & $b_{6}+s_8$\\
 $c_3+s_{18}$ & $c_{6}+s_9$\\
 $d_3+s_{23}$ & $d_{6}+s_{29}$\\
 $e_3+s_{24}$ & $e_{6}+s_{30}$\\
 \hline
 $a_{7}+b_{4}+s_{34}$ & $a_{10}+b_1+s_{31}$ \\
 $a_{4}+c_{7}+s_{35}$ & $a_{1}+c_{10}+s_{32}$ \\ 
 $a_{8}+d_{4}+s_{25}$ & $a_{11}+d_1+s_{19}$ \\
 $a_{9}+e_{4}+s_{26}$ & $a_{12}+e_1+s_{20}$ \\
 $b_{7}+c_{4}+s_{36}$ & $b_{10}+c_1+s_{33}$ \\
 $b_{8}+d_{5}+s_{27}$ & $b_{11}+d_2+s_{21}$ \\ 
 $b_{9}+e_{5}+s_{28}$ & $b_{12}+e_2+s_{22}$ \\
 $c_{8}+d_{6}+s_{29}$ & $c_{11}+d_3+s_{23}$ \\
 $c_{9}+e_{6}+s_{30}$ & $c_{12}+e_3+s_{24}$ \\
 $d_{7}+e_{7}+s_{37}$ & $d_{8}+e_{8}+s_{38}$ \\
 \hline
 $a_{13}+b_{5}+c_{5}+d_{8}+e_{8}+s_{38}$ & $a_{2}+b_{13}+c_{2}+d_{7}+e_{7}+s_{37}$ \\
 \hline
\end{tabular}
\end{center}
\vspace*{-0.4cm}
\caption{The query table for the case $K = 5, P = 3, N = 2$.}
\label{table6.2} 
\end{table}

In round $2$, for every $2$-sum containing only one desired message symbol, we add a side information symbol downloaded from the other database which already contains a common randomness symbol; for instance, we add $d_4+s_{25}$ that is already downloaded from database $2$, to the desired bit $a_8$ at database $1$, i.e., we download $a_8+d_4+s_{25}$. On the other hand, for every $2$-sum containing two of the desired message symbols, we add a new distinct common randomness symbol and download it separately from the other database; for instance, for the download $a_7+b_4$ from database $1$, we add $s_{34}$ and download $s_{34}$ separately from database $2$, and download $a_7+b_4+s_{34}$. Therefore, for this, we need to download $3$ common randomness symbols ($s_{34}, s_{35}, s_{36})$ from database $2$. Further, for every $2$-sum not containing any desired message symbols, we add a new distinct common randomness symbol with a cumulatively increasing index; for instance, for the download $d_7+e_7$ from database $1$, we add $s_{37}$ which is a new non-downloaded common randomness symbol, and download $b_7+c_7+s_{37}$. We skip rounds $3$ and $4$ because $\alpha_3 = \alpha_4 = 0$. Finally, in round $5$, where we download $5$-sums, we add the side information generated at the other databases; for instance, we add $d_8+e_8+s_{38}$ downloaded from database $2$, to $a_{13}+b_{5}+c_{5}$ and download $a_{13}+b_{5}+c_{5}+d_8+e_8+s_{38}$. This completes the achievable scheme for this case. The complete query table is shown in Table~\ref{table6.2}.

Now, we calculate the rate of this scheme. We downloaded $13$ $a$ symbols, $13$ $b$ symbols and $12$ $c$ symbols, hence a total of $L=38$ desired symbols. The total number of downloads is $D=76$. Thus, the rate $R$ of this scheme is $\frac{38}{76}=\frac{1}{2}=1-\frac{1}{2}$, which matches the capacity expression. In addition, we used $38$ common randomness symbols, hence the required common randomness $H(S)$ is $38=\frac{38}{1}$, which matches the required minimum.

We finally note that, since we downloaded asymmetric number of symbols from desired messages, i.e., $13$ $a$ symbols, $13$ $b$ symbols and $12$ $c$ symbols, we can repeat this scheme $3$ times changing the roles of $a$, $b$ and $c$, and have a symmetric scheme where we download $38$ $a$ symbols, $38$ $b$ symbols and $38$ $c$ symbols. This will not change the normalized download cost and normalized downloaded common randomness symbol numbers, hence, all the calculations (rate and common randomness calculations) will remain the same.
\end{example}

\subsubsection{General Achievable Scheme} \label{achi_MMSPIR}
Our achievability scheme is primarily based on the one in \cite{MM-PIR}, with the addition of downloading and/or mixing common randomness variables into symbol downloads appropriately. We note that, here we extend the \emph{near-optimal} algorithm in \cite{MM-PIR}, which was originally proposed for $P \leq \frac{K}{2}$, to the case of $P \geq \frac{K}{2}$, and therefore, use it for all $1 \leq P \leq K-1$ (see Footnote~\ref{footnote6}). Our achievability scheme comprises the following steps:

\begin{enumerate}
\item \emph{Initial MM-PIR Query Generation:} Generate an initial query table strictly following the near-optimal procedure in \cite{MM-PIR} for arbitrary $K$, $P$ and $N$. \label{step1} 
    
\item \emph{Repetition:} Repeat Step~\ref{step1} for a total of $\nu$ times. The purpose of the repetition is to \emph{i)} get an integer number of common randomness generated at each database by a symmetric algorithm (as exemplified in Example~\ref{example3}), and \emph{ii)} get equal number of symbols downloaded from each desired message (as exemplified in Example~\ref{example4}). Let $\nu_0$ be the smallest integer such that $\frac{(N-1)^{K-P} N \nu_0}{P}$ (i.e., $\frac{\alpha_K N \nu_0}{P}$) is an integer. Similarly, for $1 \leq k \leq \min\{P,K-P\}$, let $\nu_k$ be the smallest integer such that $\frac{\binom{P}{k} \alpha_k \nu_k }{N-1}$ is an integer ($k \leq K-P$ comes from $\alpha_{K-P+1} = \cdots = \alpha_{P-1} = 0$ in \cite[eqn.~(51)]{MM-PIR}). Then, choose $\nu$ as the lowest common multiple of these $\nu_k$, where $k \in [0:\min\{P,K-P\}]$. \label{step2}
    
\item \emph{Common Randomness Assignment:} Assign the common randomness as follows:
    \begin{enumerate}
    \item In round $1$, assign $\frac{\nu P \alpha_1}{N-1}$ independent common randomness symbols to each database, and download them. At each database, mix every $1$-sum symbol containing a desired message symbol with an arbitrary common randomness symbol already downloaded from another database, making sure that every $1$-sum symbol at each database is mixed with a different common randomness symbol. Mix all other $1$-sum symbols not containing a desired symbol with a new common randomness symbol which is not downloaded by the user. \label{step3.1}
    
    \item In round $k$ ($k\geq 2$), assign $\frac{\nu\binom{P}{k} \alpha_k}{N-1}$ independent common randomness symbols to each database, and download them. At each database: Mix every $k$-sum symbol containing only desired message symbols with an arbitrary common randomness symbol already downloaded from another database. Mix every $k$-sum symbol containing $p$ desired message symbols ($1 \leq p \leq k-1$) with the common randomness symbol from the $(k-p)$-sum symbol having the same $k-p$ undesired message symbols downloaded at any other database. Mix every $k$-sum symbol not containing any desired message symbols with a new common randomness symbol which is not downloaded by the user. \label{step3.2}
    
\item Repeat Step~\ref{step3.2} until $k$ reaches $K$. Note that if $\alpha_k=0$, nothing is done. 
\end{enumerate}
\label{step3}
\end{enumerate}

This scheme inherits the user privacy  property from the underlying scheme in \cite{MM-PIR}, as the new common randomness symbols, which are separately downloaded and subtracted out, make no difference. Due to the procedure in Step~\ref{step3}, where non-downloaded common randomness symbols are added to the downloads, no undesired symbol is decodable because of the added unknown common randomness, ensuring the database privacy constraint.

\subsubsection{Rate and Common Randomness Amount Calculation} \label{rate-calculation}
We calculate the achievable rate and the minimal required common randomness for only one repetition of the scheme. The reason for this is that, in every repetition, every involved term would be multiplied by $T$, and thus $T$ can be cancelled in the numerator and the denominator of the normalized rate and normalized required common randomness expressions. 

For each database, before the assignment of common randomness, let $D_1$ be the total number of downloaded symbols, $U_1$ be the total number of downloaded undesired symbols, $U_2$ be the total number of downloaded symbols including only desired message symbols, and $D_2$ be the total number of downloaded common randomness symbols. The achievable rate is then given by,
\begin{align}
R = \frac{D_1-U_1}{D_1+D_2}
\end{align}

Using the respective results in \cite[eqns.~(66)-(69) and (70)-(72)]{MM-PIR}, we have
\begin{align}
 D_1 &= \sum_{k=1}^K \binom{K}{k}\alpha_k = \sum_{i=1}^P \gamma_i r_i^{K-P}\left[\left(1+\frac{1}{r_i}\right)^K-1\right]\\
 U_1 &= \sum_{k=1}^{K-P} \binom{K-P}{k}\alpha_k = \sum_{i=1}^P \gamma_i r_i^{K-P}\left[\left(1+\frac{1}{r_i}\right)^{K-P}-1\right]
\end{align}
In the proposed new achievable scheme, every $k$-sum symbol ($1 \leq k \leq \min\{P,K-P\}$) containing only desired message symbols is mixed with an arbitrary common randomness symbol which is downloaded from another database. In addition, these downloaded common randomness symbols are uniformly requested from the other $(N-1)$ databases. Thus,
\begin{align}
 U_2 &= \sum_{k=1}^{\min\{K-P,P\}} \binom{P}{k}\alpha_k \\
 D_2 &= \frac{1}{N-1} U_2 = \frac{1}{N-1} \sum_{k=1}^{\min\{K-P,P\}} \binom{P}{k}\alpha_k    
\end{align}

With these observations we have the following two lemmas where we compute the MM-SPIR rate and the required common randomness amount.

\begin{lemma} \label{lemma8}
The rate of the proposed achievable scheme is,
\begin{align}
 R = 1-\frac{1}{N}
\end{align}
\end{lemma}

\begin{Proof} We first calculate $D_2$ in two possible settings. When $P \leq \frac{K}{2}$, i.e., $P \leq K-P$, 
\begin{align}
D_2 &= \frac{1}{N-1} \sum_{k=1}^{P} \binom{P}{k}\alpha_k \\
    &= \frac{1}{N-1} \sum_{k=1}^{P} \binom{P}{k} \sum_{i=1}^{P} \gamma_i r_i^{K-P-k} \\
    &= \frac{1}{N-1} \sum_{k=1}^{P} \sum_{i=1}^{P} \binom{P}{k} \gamma_i r_i^{K-P-k} \\
    &= \frac{1}{N-1} \sum_{i=1}^{P} \sum_{k=1}^{P} \binom{P}{k} \gamma_i r_i^{K-P-k} \\
    &= \frac{1}{N-1} \sum_{i=1}^{P} \gamma_i r_i^{K-2P} \sum_{k=1}^{P} \binom{P}{k} r_i^{P-k} \\
    &= \frac{1}{N-1} \sum_{i=1}^{P} \gamma_i r_i^{K-2P} (N-1) r_i^P \label{eq8.1} \\
    &= \frac{1}{N-1} \sum_{i=1}^{P} \gamma_i r_i^{K-P} (N-1) \label{eq8.2}
\end{align}
where \eqref{eq8.1} follows because $r_i$ is a root of the characteristic equation \cite[eqn.~(59)]{MM-PIR}.

When $\frac{K}{2} \leq P \leq K-1$, i.e., $K-P \leq P$,
\begin{align}
D_2 &= \frac{1}{N-1} \sum_{k=1}^{K-P} \binom{P}{k}\alpha_k \\
    &= \frac{1}{N-1} \sum_{k=1}^{P} \binom{P}{k}\alpha_k - \sum_{k=K-P+1}^{P} \binom{P}{k}\alpha_k \\
    &= \frac{1}{N-1} \sum_{k=1}^{P} \binom{P}{k}\alpha_k \label{eq8.3} \\
    &= \frac{1}{N-1} \sum_{i=1}^{P} \gamma_i r_i^{K-P} (N-1) \label{eq8.4}
\end{align}
where \eqref{eq8.3} follows because $\alpha_{K-P+1} = \cdots = \alpha_{P-1} = 0$ due to \cite[eqn.~(51)]{MM-PIR}, and \eqref{eq8.4} follows from \eqref{eq8.2}.

Therefore, from (\ref{eq8.2}) and (\ref{eq8.4}), for all $P$, where $1 \leq P \leq K-1$, we always have
\begin{align}
D_2 = \frac{1}{N-1} \sum\limits_{k=1}^{P} \binom{P}{k}\alpha_k = \frac{1}{N-1} \sum\limits_{i=1}^{P} \gamma_i r_i^{K-P} (N-1) 
\end{align} 

Now, in order to show that $R = \frac{D_1-U_1}{D_1+D_2} = 1 - \frac{1}{N}$, we need to equivalently show that $D_1 = NU_1 + (N-1)D_2$. Thus, we proceed as,
\begin{align}
NU_1 + (N-1)D_2 
&= N \sum_{i=1}^P \gamma_i r_i^{K-P}\left[\left(1+\frac{1}{r_i}\right)^{K-P}-1\right]
+ \sum_{i=1}^{P} \gamma_i r_i^{K-P} (N-1) \\
&= \sum_{i=1}^P \gamma_i r_i^{K-P}\left[N\left(1+\frac{1}{r_i}\right)^{K-P}-N+N-1\right] \\
&= \sum_{i=1}^P \gamma_i r_i^{K-P}\left[N\left(1+\frac{1}{r_i}\right)^{K-P}-1\right] \\
&= \sum_{i=1}^P \gamma_i r_i^{K-P}\left[N\left(1+\frac{1}{r_i}\right)^{-P}\left(1+\frac{1}{r_i}\right)^{K}-1\right] \\
&= \sum_{i=1}^P \gamma_i r_i^{K-P}\left[\left(1+\frac{1}{r_i}\right)^{K}-1\right] \label{eq8.5} \\
&= D_1 \label{eq8.6}
\end{align}
where \eqref{eq8.5} follows because $N(1+\frac{1}{r_i})^{-P} = 1$, which comes from \cite[eqn.~(62)]{MM-PIR}.
\end{Proof}

\begin{lemma} \label{lemma9}
The minimal required common randomness in the proposed achievable scheme is,
\begin{align}
H(S) = \frac{PL}{N-1} 
\end{align}
\end{lemma}

\begin{Proof}
In our proposed scheme, at each database, a new common randomness symbol is employed only in two cases. The first case is when a new common randomness symbol is added to a $k$-sum symbol that contains only desired message symbols. In this case, the common randomness symbols are equally distributed over the $(N-1)$ databases and downloaded from them. The second case is when a new common randomness symbol is assigned to a $k$-sum symbol that does not contain any desired message symbol. In this case, the common randomness symbols are not downloaded. Therefore, we count the total number of distinct common randomness symbols as $H(S) = U_1 + D_2$. We note that $L$ can be written as $\frac{1}{P}(D_1 - U_1)$. Thus,
\begin{align}
\frac{PL}{N-1} &= \frac{\frac{P}{P}(D_1 - U_1)}{N-1} \\
               &= \frac{D_1 - U_1}{N-1} \\
               &= \frac{NU_1 + (N-1)D_2 - U_1}{N-1} \label{eq9.1} \\
               &= \frac{(N-1)U_1 + (N-1)D_2}{N-1} \\
               &= U_1 + D_2 \\
               &= H(S)
\end{align}
where \eqref{eq9.1} comes from \eqref{eq8.6}, i.e., $D_1 = NU_1 + (N-1)D_2 $.
\end{Proof}

\section{MM-LSPIR: Arbitrary Message Lengths} \label{MM-LSPIR}
Since the message sizes in the PSI problem are given and fixed, in particular, they are fixed to be $1$ (as the incidence vectors are composed $0$s and $1$s), we need to determine the capacity of MM-SPIR with a given and fixed message size $L$. We call this setting MM-LSPIR. The capacity of MM-LSPIR is given in the next theorem.

\begin{theorem}\label{Thm3}
The MM-LSPIR capacity for $N \geq 2$, $K \geq 2$, and $P\leq K$, for an arbitrary message length $L$ is given by,
\begin{align}
C_{MM-LSPIR} = 
 \begin{cases}
 1, & P=K\\
 \frac{PL}{\left\lceil\frac{NPL}{N-1}\right\rceil}, & 1 \leq P \leq K-1, ~ H(S) \geq \left\lceil\frac{PL}{N-1}\right\rceil \\
 0, & \text{otherwise}
 \end{cases}
\end{align}
\end{theorem}

We give the converse of Theorem~\ref{Thm3} in Section~\ref{MM-LSPIR-conv}, the achievability in Section~\ref{MM-LSPIR-ach}, and map MM-LSPIR back to PSI in Section~\ref{MM-LSPIR-map-back}.

\subsection{MM-LSPIR: Converse Proof} \label{MM-LSPIR-conv}
From the converse proof of Theorem~\ref{Thm2}, using \eqref{ratedefinition} and \eqref{rateresult}, we have
\begin{align}
  R_{MM-LSPIR} = \frac{PL}{D_{MM-LSPIR}} \leq \frac{PL}{\sum_{n=1}^N  H(A_n^{[\mathcal{P}]})} \leq \frac{N-1}{N} = 1 - \frac{1}{N}  
\end{align}

Note that, for an arbitrary finite fixed message length $L$, the download cost $D_{MM-LSPIR}$ must be a positive integer. Thus, we have,
\begin{align}
  D_{MM-LSPIR} \geq \left\lceil\frac{NPL}{N-1}\right\rceil
\end{align}
and therefore, the converse result for a finite and fixed $L$, is
\begin{align}
	R_{MM-LSPIR} = \frac{PL}{D_{MM-LSPIR}} \leq \frac{PL}{\left\lceil\frac{NPL}{N-1}\right\rceil} \label{ms-lspir-conv1} 
\end{align}

Similarly, the entropy of common randomness must also be a positive integer, as the common randomness symbols are picked uniformly and independently from the same field as the message symbols. Thus, with a careful look at going from \eqref{cr1.8} to \eqref{crresult}, we have,
\begin{align}
  H(S) \geq \left\lceil\frac{PL}{N-1}\right\rceil \label{ms-lspir-conv2}  
\end{align}
Therefore, \eqref{ms-lspir-conv1} and \eqref{ms-lspir-conv2} constitute the converse for Theorem~\ref{Thm3}.

\subsection{MM-LSPIR: Achievability Proof} \label{MM-LSPIR-ach}
For simplicity, we follow the achievability scheme in \cite[Section~IV.B.1]{SPIR}. By setting the value of $l_K$ to be $1$ and using the total length of multi-messages $PL$ to replace the length of a single message $L$, we get the achievability of MM-LSPIR directly with $D = \left\lceil\frac{NPL}{N-1}\right\rceil$ and $H(S) = \left\lceil\frac{PL}{N-1}\right\rceil$. These constitute the achievability for Theorem~\ref{Thm3}. The achievability can also be done by using an extension of our proposed alternative achievable scheme.  

\subsection{Mapping MM-LSPIR Back to PSI} \label{MM-LSPIR-map-back}
Finally, we map our MM-SPIR results back to the PSI problem to obtain Theorem~\ref{Thm1}. Recall that, in the PSI problem, by generating the sets $\cp_1$ and $\cp_2$ by i.i.d.~drawing the elements from the alphabet $\cp_{alph}$, we obtain i.i.d.~messages in the corresponding MM-SPIR problem. Further, by choosing the probability $q_i$ of choosing each element to be included in the set $\cp_i$ to be $q_i=\frac{1}{2}$, for $i=1, 2$, we obtain uniformly distributed messages, with message size $L=1$. Therefore, the PSI problem is equivalent to an MM-LSPIR problem with $L=1$. Now, using Theorem~\ref{Thm3} with $L=1$, we obtain the ultimate result of this paper in Theorem~\ref{Thm1}. 

\section{Conclusion and Discussion}
We investigated the PSI problem over a finite set $\mathbb{S}_K$ from an information-theoretic point of view. We showed that the problem can be recast as an MM-SPIR problem with a message size $1$. This is under the assumption that the sets (or their corresponding incidence vectors) can be stored in replicated and non-colluding databases. Further, the set elements are generated in an i.i.d.~fashion with a probability $\frac{1}{2}$ of adding any element to any of the sets. 

To that end, we explored the information-theoretic capacity of MM-SPIR as a stand-alone problem. We showed that joint multi-message retrieval does not outperform the successive application of single-message SPIR. This is unlike the case of MM-PIR, where significant performance gains can be obtained due to joint retrieval. We remark that SM-SPIR is a special case of the problem studied in this paper by plugging $P=1$. For the converse proof, we extended the proof techniques of \cite{SPIR} to the setting of multi-messages. In particular, the proof of Lemma~\ref{lemma6} is significantly more involved than the proof in \cite{SPIR}. This is due to the fact that the desired message subsets in the case of MM-SPIR may not be disjoint. To unify the query structures of MM-PIR and MM-SPIR, we proposed a new capacity-achieving scheme for any $P$ as an alternative to the successive usage of the scheme in \cite{SPIR}. Our scheme primarily consists of three steps: Exploiting the achievable scheme in \cite{MM-PIR}, making necessary repetitions to symmetrize the scheme, and adding the needed common randomness properly. The last step is inspired by \cite{SPIR_Mismatched}. Based on these results, we showed that the optimal download cost for PSI is $\min\left\{\left\lceil\frac{P_1 N_2}{N_2-1}\right\rceil, \left\lceil\frac{P_2 N_1}{N_1-1}\right\rceil\right\}$. 

In the following subsections, we make a few remarks about assumptions made in this paper, and directions for further research.

\subsection{Data Generation Model}
In this work, we add elements to each set in an i.i.d.~manner and with probability $\frac{1}{2}$. This assumption is made for two reasons, first, to have i.i.d.~incidence vectors, therefore, i.i.d.~messages in the MM-SPIR problem, and second, to have uniform messages to avoid the need for compressing the messages $W_{1:K}$ before/within retrieval. However, this assumption may be restrictive, as with this assumption, the expected sizes of both sets are $\frac{K}{2}$. Even with keeping the i.i.d.~generation assumption, the probability of adding each element to set $i$ could be generalized to be an arbitrary $q_i$. In this more general case, the expected sizes of the sets, $Kq_1$ and $Kq_2$, could be arbitrary. This may be done by using appropriate compression before/during retrieval, but needs to be studied further. Regarding the i.i.d.~selection of elements, while this assumption is not needed from the achievability side, it is needed for the converse proof. To overcome these restrictions, as future work, it may be worthwhile to investigate the MM-SPIR problem with correlated messages.

\subsection{Upload Cost Reduction}\label{upload}
In this paper, we have focused on the download cost as the sole performance metric. A more natural performance metric is to consider the combined upload and download cost. In this section, we provide an illustrative example, which shows that the upload cost may be reduced without sacrificing the download cost. Nevertheless, the characterization of the optimal combined upload and download cost  is an interesting future direction that is outside the scope of this paper.
\begin{example}
    Consider the SPIR problem with $K = 3$, $N = 2$, $P = 1$, $L =1$. The original SPIR scheme in \cite{SPIR} achieves the optimal download cost of $D = 2$ bits, while the upload cost is $U = 6$ bits. Inspired by \cite{Min_Uploadcost_SPIR}, we show that the upload cost can be reduced to just $4$ bits without increasing the download cost. Our new achievable scheme is as follows:  
    
    For any one of the two databases, there are four possible answers $A_n^{(q)}$, where $n \in [2], q \in [4]$ and common randomness $S$ is a uniformly distributed bit:
\begin{align}
    A_1^{(1)} &= W_1+W_2+W_3+S, \qquad A_2^{(1)} = W_2+W_3+S \\
    A_1^{(2)} &= W_1+S, \qquad \qquad \qquad \quad A_2^{(2)} = S \\
    A_1^{(3)} &= W_2+S, \qquad \qquad \qquad \quad A_2^{(3)} = W_1+W_2+S \\
    A_1^{(4)} &= W_3+S, \qquad \qquad \qquad \quad A_2^{(4)} = W_1+W_3+S 
\end{align}

The corresponding queries for different desired messages are generated according to the following distributions:
\begin{align}
    \quad W_1: (Q_1^{[1]},Q_2^{[1]}) \quad \text{is uniform over} \quad \{(1,1),(2,2),(3,3),(4,4)\}, \\
    \quad W_2: (Q_1^{[2]},Q_2^{[2]}) \quad \text{is uniform over} \quad \{(1,4),(2,3),(3,2),(4,1)\}, \\
    \quad W_3: (Q_1^{[3]},Q_2^{[3]}) \quad \text{is uniform over} \quad \{(1,3),(2,4),(3,1),(4,2)\}.
\end{align}

The reliability constraint follows from the fact for every query pair, the user can cancel the interfering messages and the common randomness $S$ from the other database. For the database-privacy constraint, we note that the undesired messages are always mixed with $S$. Hence, the information leakage from undesired messages is zero. For the user-privacy constraint, we have
\begin{align}
    P(Q_n^{[k]}=q) = P(Q_n^{[k^\prime]}=q), \quad \forall k, k^\prime \in [3], \forall n \in [2], \forall q \in [4]
\end{align}
i.e., from the point of view of any database, the same set of queries is used for any desired message $W_i$, where $i=1,2,3$ with the same probability distribution.

For the proposed scheme, the required download cost is $D = 2$ bits and the required upload cost is $U = 4$ bits, which outperforms the one in \cite{SPIR} in terms of upload cost.
\end{example}

\subsection{Communication Model}    
We note that our optimality result is restricted to the presented communication scenario, where a sender submits queries to a receiver in one round. An interesting future direction is to investigate whether there is a more efficient communication scheme or whether there can be an impossibility result that can assert that no other communication scheme can outperform our presented scheme. 

\subsection{Single Database Assumption}
Our scheme is infeasible for $N_1=N_2=1$ due to the capacity result for MM-SPIR. It would be interesting to see if MM-SPIR can be made feasible with certain modifications to the problem, e.g., side information, or alternatively, if PSI can be transformed into other problems, in the case of a single-server.   

\bibliographystyle{unsrt}
\bibliography{PSI}

\end{document}